\documentclass[11pt]{article}

\usepackage{graphics}
\usepackage{feynarts}
\usepackage{cite}
\usepackage{axodraw}
\usepackage{rotating}
\usepackage{url}

\makeatletter
\if@twoside \oddsidemargin -17pt \evensidemargin 00pt
\else \oddsidemargin 00pt \evensidemargin 00pt
\fi
\expandafter\ifx\csname mathrm\endcsname\relax\def\mathrm#1{{\rm #1}}\fi

\renewcommand{\theequation}{\thesection.\arabic{equation}}
\newcounter{saveeqn}

\@addtoreset{equation}{section}

\makeatother

\def\beq{\begin{equation}}
\def\eeq{\end{equation}}
\def\beqar{\begin{eqnarray}}
\def\eeqar{\end{eqnarray}}
\def\barr#1{\begin{array}{#1}}
\def\earr{\end{array}}
\def\bfi{\begin{figure}}
\def\efi{\end{figure}}
\def\btab{\begin{table}}
\def\etab{\end{table}}
\def\bce{\begin{center}}
\def\ece{\end{center}}

\def\text{\textstyle}

\def\al{\alpha}

\def\ga{\gamma}

\def\si{\sigma}
\def\Ga{\Gamma}
\def\De{\Delta}

\def\refeq#1{\mbox{(\ref{#1})}}
\def\reffi#1{\mbox{Fig.~\ref{#1}}}

\def\refta#1{\mbox{Table~\ref{#1}}}

\def\refse#1{\mbox{Section~\ref{#1}}}

\def\citere#1{\mbox{Ref.~\cite{#1}}}
\def\citeres#1{\mbox{Refs.~\cite{#1}}}

\newcommand{\GeV}{\unskip\,\mathrm{GeV}}
\newcommand{\MeV}{\unskip\,\mathrm{MeV}}
\newcommand{\TeV}{\unskip\,\mathrm{TeV}}
\newcommand{\fba}{\unskip\,\mathrm{fb}}

\def\mathswitch#1{\relax\ifmmode#1\else$#1$\fi}
\def\mathswitchr#1{\relax\ifmmode{\mathrm{#1}}\else$\mathrm{#1}$\fi}
\def\mathswitchit#1{\relax\ifmmode{#1}\else$#1$\fi}

\newcommand{\PV}{\mathswitch V}
\newcommand{\PW}{\mathswitchr W}
\newcommand{\PZ}{\mathswitchr Z}

\newcommand{\PH}{\mathswitchr H}

\newcommand{\Pe}{\mathswitchr e}

\newcommand{\Pd}{\mathswitchr d}

\newcommand{\Pl}{\mathswitch l}

\newcommand{\Pu}{\mathswitchr u}
\newcommand{\Ps}{\mathswitchr s}
\newcommand{\Pb}{\mathswitchr b}
\newcommand{\Pc}{\mathswitchr c}
\newcommand{\Pt}{\mathswitchr t}
\newcommand{\Pp}{\mathswitchr p}

\newcommand{\MV}{\mathswitch {M_\PV}}
\newcommand{\MW}{\mathswitch {M_\PW}}

\newcommand{\MZ}{\mathswitch {M_\PZ}}
\newcommand{\MH}{\mathswitch {M_\PH}}

\newcommand{\Mt}{\mathswitch {m_\Pt}}
\newcommand{\GV}{\mathswitch {\Gamma_\PV}}
\newcommand{\GW}{\mathswitch {\Gamma_\PW}}
\newcommand{\GZ}{\mathswitch {\Gamma_\PZ}}

\newcommand{\GF}{\mathswitch {G_\mu}}

\def\ie{i.e.\ }
\def\eg{e.g.\ }

\newcommand{\ri}{{\mathrm{i}}}

\newcommand{\EW}{{\mathrm{EW}}}
\newcommand{\QCD}{{\mathrm{QCD}}}

\newcommand{\OS}{{\mathrm{OS}}}

\newcommand{\NLO}{{\mathrm{NLO}}}

\def\draftdate{\relax}
\def\mda{\relax}
\def\mua{\relax}
\def\mla{\relax}
\def\draft{
\def\thtystars{******************************}
\def\sixtystars{\thtystars\thtystars}
\typeout{}
\typeout{\sixtystars**}
\typeout{* Draft mode!
         For final version remove \protect\draft\space in source file *}
\typeout{\sixtystars**}
\typeout{}
\def\draftdate{\today}
\def\mua{\marginpar[\boldmath\hfil$\uparrow$]%
                   {\boldmath$\uparrow$\hfil}%
                    \typeout{marginpar: $\uparrow$}\ignorespaces}
\def\mda{\marginpar[\boldmath\hfil$\downarrow$]%
                   {\boldmath$\downarrow$\hfil}%
                    \typeout{marginpar: $\downarrow$}\ignorespaces}
\def\mla{\marginpar[\boldmath\hfil$\rightarrow$]%
                   {\boldmath$\leftarrow $\hfil}%
                    \typeout{marginpar: $\leftrightarrow$}\ignorespaces}
\def\Mua{\marginpar[\boldmath\hfil$\Uparrow$]%
                   {\boldmath$\Uparrow$\hfil}%
                    \typeout{marginpar: $\Uparrow$}\ignorespaces}
\def\Mda{\marginpar[\boldmath\hfil$\Downarrow$]%
                   {\boldmath$\Downarrow$\hfil}%
                    \typeout{marginpar: $\Downarrow$}\ignorespaces}
\def\Mla{\marginpar[\boldmath\hfil$\Rightarrow$]%
                   {\boldmath$\Leftarrow $\hfil}%
                    \typeout{marginpar: $\Leftrightarrow$}\ignorespaces}
\overfullrule 5pt
\oddsidemargin -15mm
\marginparwidth 29mm
}

\makeatletter

\def\eqnarray{\stepcounter{equation}\let\@currentlabel=\theequation
\global\@eqnswtrue
\global\@eqcnt\z@\tabskip\@centering\let\\=\@eqncr
$$\halign to \displaywidth\bgroup\hskip\@centering
  $\displaystyle\tabskip\z@{##}$\@eqnsel&\global\@eqcnt\@ne
  \hskip 2\arraycolsep \hfil${##}$\hfil
  &\global\@eqcnt\tw@ \hskip 2\arraycolsep $\displaystyle\tabskip\z@{##}$\hfil
   \tabskip\@centering&\llap{##}\tabskip\z@\cr}
\def\appendix{\par
 \setcounter{section}{0} \setcounter{subsection}{0}
 \def\thesection{\Alph{section}}}

\newcommand{\lsim}
{\;\raisebox{-.3em}{$\stackrel{\displaystyle <}{\sim}$}\;}
\newcommand{\gsim}
{\;\raisebox{-.3em}{$\stackrel{\displaystyle >}{\sim}$}\;}

\def\dsl{\mathpalette\make@slash}
\def\make@slash#1#2{\setbox\z@\hbox{$#1#2$}%
  \hbox to 0pt{\hss$#1/$\hss\kern-\wd0}\box0}

\newcommand{\mycaption}[2][]{{\begin{center} \parbox{15cm}{{\bf \caption[#1]{\rm {#2}}}} \end{center} }}

\makeatother

\newcommand{\phz}{\phantom{(0)}}

\newcommand{\processes}{$\Pp\Pp/\Pp\bar\Pp\to\PH + l\nu_l/l^-l^+/\nu_l\bar\nu_l+X$}
\newcommand{\processesres}{$\Pp\Pp/\Pp\bar\Pp\to \PH+\PW/\PZ \to\PH + l\nu_l/l^-l^+/\nu_l\bar\nu_l+X$}
\newcommand{\finalstates}{$\PH + l\nu_l/l^-l^+/\nu_l\bar\nu_l+X$}

\begin{document}

\thispagestyle{empty}
\def\thefootnote{\fnsymbol{footnote}}
\setcounter{footnote}{1}
\null
\draftdate\hfill  
FR-PHENO-2011-025 \\
\strut\hfill PSI-PR-11-04 \\
\strut\hfill ZU-TH 29/11 \\
\strut\hfill TTK-11-61 \\
\vskip 0cm
\vfill
\begin{center}
  {\Large \boldmath{\bf Electroweak corrections
            to Higgs-strahlung off W/Z bosons\\[.5em] 
            at the Tevatron and the LHC with {\sc Hawk}
      }
\par} \vskip 2.5em
{\large
{\sc Ansgar Denner$^{1}$, Stefan Dittmaier$^{2}$, 
Stefan Kallweit$^{3,4}$, \\[.3em] and Alexander M\"uck$^{5}$
}\\[1ex]
{\normalsize \it 
$^1$Universit\"at W\"urzburg, Institut f\"ur Theoretische Physik und Astrophysik, \\
D-97074 W\"urzburg, Germany
}\\[1ex]
{\normalsize \it 
$^2$Albert-Ludwigs-Universit\"at Freiburg, Physikalisches Institut, \\
D-79104 Freiburg, Germany
}\\[1ex]
{\normalsize \it 
$^3$Paul Scherrer Institut, W\"urenlingen und Villigen,\\ 
CH-5232 Villigen PSI, Switzerland
}\\[1ex]
{\normalsize \it 
$^4$Institut f\"ur Theoretische Physik, Universit\"at Z\"urich,\\
CH-8057 Z\"urich, Switzerland
}\\[1ex]
{\normalsize \it 
$^5$Institut f\"ur Theoretische Teilchenphysik und Kosmologie, RWTH Aachen University, \\
D-52056 Aachen, Germany}
\\[2ex]
}
\par \vskip 1em
\end{center}\par
\vskip .0cm \vfill {\bf Abstract:} \par 
The associate production of
Higgs bosons with W or Z bosons, known as Higgs-strahlung,
is an important search channel for Higgs bosons at the hadron colliders
Tevatron and LHC for low Higgs-boson masses.  We refine a previous
calculation of next-to-leading-order electroweak corrections (and
recalculate the QCD corrections) upon including the leptonic decay of
the W/Z bosons, thereby keeping the fully differential information of
the 2-lepton + Higgs final state.  The gauge invariance of the
W/Z-resonance treatment is ensured by the use of the complex-mass
scheme.  The electroweak corrections, which are at the level of
$-(5{-}10)\%$ for total cross sections, further increase in size with
increasing transverse momenta $p_{\mathrm{T}}$ in differential cross
sections. For instance, for $p_{\mathrm{T,H}}\gsim200\GeV$, which is
the interesting range at the LHC, the electroweak corrections to WH
production reach about $-14\%$ for $\MH=120\GeV$.
The described corrections are implemented in the {\sc Hawk} Monte Carlo
program, which was initially designed for the vector-boson-fusion channel,
and are discussed for various distributions in the production channels \processes.
\par
\vskip 1cm
\noindent
December 2011
\par
\null
\setcounter{page}{0}
\clearpage
\def\thefootnote{\arabic{footnote}}
\setcounter{footnote}{0}

\section{Introduction}

The Higgs-strahlung processes at hadron colliders, \processesres,
represent important search channels for a Standard Model (SM) Higgs
boson in the low-mass range, which is not only favoured by electroweak
precision tests, but also by recent results from Higgs searches at the
LHC \cite{Higgs_atlas_cms}. At the Tevatron the sensitivity to a light
SM Higgs boson almost exclusively results from these channels, and the
analysis of the full Tevatron data set might be able to exclude values
of the Higgs-boson mass $\MH$ close to the LEP exclusion limit of
$\MH=114\GeV$ beyond the 95\% confidence level.  At the LHC this
low-mass range is explored mainly via inclusive Higgs production
employing the $\PH\to\ga\ga$ decay channel, but Higgs-strahlung
slightly improves the sensitivity and provides an interesting
additional independent search channel with the Higgs boson decaying
via $\PH\to\Pb\bar\Pb$.  The challenge in all these analyses is
background control, both in cross-section normalization and
differential distributions.  This is particularly pronounced in the
boosted $\PH+\PW/\PZ$ analysis at the LHC where the $\PH\to\Pb\bar\Pb$
decay should be identified from the b-quark substructure within a
highly boosted ``fat jet''~\cite{Butterworth:2008iy}.

The mentioned analyses call for precise theoretical predictions with
proper estimates of the corresponding theoretical and parametric
uncertainties.  The next-to-leading-order (NLO) QCD corrections, whose
structure is identical to the respective corrections to the Drell--Yan
process, were calculated~\cite{Han:1991ia} a long time ago and are
available, e.g., in the public programs {\sc VV2H}~\cite{vv2h} and
{\sc MCFM}~\cite{mcfm}.  The most important part of the
next-to-next-to-leading-order (NNLO) QCD corrections is similar to the
Drell--Yan process as well and was calculated in
\citere{Brein:2003wg}, using the results of \citere{Hamberg:1990np} on
the total Drell--Yan cross section. The production of $\PZ\PH$ final
states additionally receives contributions from gluon fusion via
top-quark loops, which was also included in the calculation of
\citere{Brein:2003wg}.  The remaining NNLO QCD corrections that are
not of Drell--Yan type, which involve Higgs couplings to top-quark
loops, have been shown to be at the $1{-}2\%$
level~\cite{arXiv:1111.0761}.  The calculation of NLO corrections,
which are of the order of $+30\%$ in QCD, was completed by the NLO
electroweak (EW) corrections~\cite{Ciccolini:2003jy}, which turned out
to be about $-(5{-}10)\%$, and a combined prediction for total cross
sections at the Tevatron and the LHC was presented in
\citere{Brein:2004ue}.  An update of the cross-section prediction,
together with a thorough estimate of uncertainties, was presented in
the first report~\cite{Dittmaier:2011ti} of the LHC Higgs Cross
Section Working Group.  For the LHC with a centre-of-mass (CM) energy
of $7(14)\TeV$ the QCD scale uncertainties were assessed to be about
$1\%$ and $1{-2}(3{-}4)\%$ for $\PW\PH$ and $\PZ\PH$ production,
respectively, while uncertainties of the parton distribution functions
(PDFs) turn out to be about $3{-}4\%$.

These results, however, hold only for total cross sections and
have to be generalized to distributions for the experimental analyses,
in particular if one employs boosted Higgs bosons.
An important step towards differential distributions has
been made recently at NNLO QCD in \citere{Ferrera:2011bk}, where the
Drell--Yan-like corrections to $\PW\PH$ production are calculated
while keeping the full kinematical information of the final state.

In this paper we present fully differential NLO EW predictions for the
final states \finalstates, i.e. including the leptonic decays of the W
and Z bosons. We also recalculate the corresponding NLO QCD
corrections.  In contrast to the (N)NLO QCD calculations the EW loop
corrections involve also irreducible corrections connecting production
and decay of the W/Z bosons. In order to guarantee the gauge
invariance of the NLO calculation in spite of the W/Z resonances we
employ the complex-mass scheme, which was introduced for leading-order
(LO) and NLO calculations in \citere{Denner:1999gp} and
\citere{Denner:2005fg}, respectively.  The use of complex W/Z masses
at the same time cures the appearance of perturbative artifacts for
Higgs-boson masses near the pair production thresholds at
$\MH=2\MW,2\MZ$, where the EW corrections to $\PW\PH/\PZ\PH$
production~\cite{Ciccolini:2003jy} are singular.  The calculation
described in this paper will be part of the Monte Carlo program {\sc
  Hawk}, which was originally designed for the description of Higgs
production via vector-boson fusion including NLO QCD and EW
corrections~\cite{Ciccolini:2007jr} and is publically
available~\cite{hawk}.

The paper is organized as follows:
In \refse{se:calc} we describe the structure of the underlying 
NLO calculation and the techniques used.
Section~\ref{se:numres} contains a detailed discussion of
numerical results for the processes \processes\
at the Tevatron and the LHC, the latter at CM energies of $7\TeV$
and $14\TeV$.
Finally, we conclude in \refse{se:conclusion}.

\section{Structure of the NLO calculation}
\label{se:calc}

\subsection{General setup}
\label{se:setup} 

At LO, the associate production of a Higgs boson $\PH$ with a weak gauge
boson $V=\PW,\PZ$ can only proceed via quark--antiquark annihilation
at hadron colliders. Treating the incoming quarks and the outgoing leptons 
as massless, the Higgs boson does not couple to the massless fermions, and 
there is only one LO diagram per channel, see \reffi{fi:borndiag}.
\bfi
\begin{center}
\begin{feynartspicture}(100,100)(1,1)

\FADiagram{}
\FAProp(0.,15.)(5.5,10.)(0.,){/Straight}{1}
\FALabel(2.18736,11.8331)[tr]{$u_i$}
\FAProp(0.,5.)(5.5,10.)(0.,){/Straight}{-1}
\FALabel(3.31264,6.83309)[tl]{$\bar d_j$}
\FAProp(20.,17.)(11.5,10.)(0.,){/ScalarDash}{0}
\FALabel(15.4036,14.0235)[br]{$\PH$}
\FAProp(20.,10.)(15.5,6.5)(0.,){/Straight}{-1}
\FALabel(17.2784,8.9935)[br]{$\nu_\Pl$}
\FAProp(20.,3.)(15.5,6.5)(0.,){/Straight}{1}
\FALabel(18.2216,5.4935)[bl]{$\Pl^+$}
\FAProp(5.5,10.)(11.5,10.)(0.,){/Sine}{0}
\FALabel(8.5,11.07)[b]{$\PW$}
\FAProp(11.5,10.)(15.5,6.5)(0.,){/Sine}{0}
\FALabel(12.9593,7.56351)[tr]{$\PW$}
\FAVert(5.5,10.){0}
\FAVert(11.5,10.){0}
\FAVert(15.5,6.5){0}

\end{feynartspicture}
\hspace*{.5em}
\begin{feynartspicture}(100,100)(1,1)

\FADiagram{}
\FAProp(0.,15.)(5.5,10.)(0.,){/Straight}{1}
\FALabel(2.18736,11.8331)[tr]{$d_j$}
\FAProp(0.,5.)(5.5,10.)(0.,){/Straight}{-1}
\FALabel(3.31264,6.83309)[tl]{$\bar u_i$}
\FAProp(20.,17.)(11.5,10.)(0.,){/ScalarDash}{0}
\FALabel(15.4036,14.0235)[br]{$\PH$}
\FAProp(20.,10.)(15.5,6.5)(0.,){/Straight}{-1}
\FALabel(17.2784,8.9935)[br]{$\Pl^-$}
\FAProp(20.,3.)(15.5,6.5)(0.,){/Straight}{1}
\FALabel(18.2216,5.4935)[bl]{$\bar\nu_\Pl$}
\FAProp(5.5,10.)(11.5,10.)(0.,){/Sine}{0}
\FALabel(8.5,11.07)[b]{$\PW$}
\FAProp(11.5,10.)(15.5,6.5)(0.,){/Sine}{0}
\FALabel(12.9593,7.56351)[tr]{$\PW$}
\FAVert(5.5,10.){0}
\FAVert(11.5,10.){0}
\FAVert(15.5,6.5){0}

\end{feynartspicture}
\hspace*{.5em}
\begin{feynartspicture}(100,100)(1,1)

\FADiagram{}
\FAProp(0.,15.)(5.5,10.)(0.,){/Straight}{1}
\FALabel(2.18736,11.8331)[tr]{$q_i$}
\FAProp(0.,5.)(5.5,10.)(0.,){/Straight}{-1}
\FALabel(3.31264,6.83309)[tl]{$\bar q_i$}
\FAProp(20.,17.)(11.5,10.)(0.,){/ScalarDash}{0}
\FALabel(15.4036,14.0235)[br]{$\PH$}
\FAProp(20.,10.)(15.5,6.5)(0.,){/Straight}{-1}
\FALabel(17.2784,8.9935)[br]{$\Pl^-$}
\FAProp(20.,3.)(15.5,6.5)(0.,){/Straight}{1}
\FALabel(18.2216,5.4935)[bl]{$\Pl^+$}
\FAProp(5.5,10.)(11.5,10.)(0.,){/Sine}{0}
\FALabel(8.5,11.07)[b]{$\PZ$}
\FAProp(11.5,10.)(15.5,6.5)(0.,){/Sine}{0}
\FALabel(12.9593,7.56351)[tr]{$\PZ$}
\FAVert(5.5,10.){0}
\FAVert(11.5,10.){0}
\FAVert(15.5,6.5){0}

\end{feynartspicture}
\hspace*{.5em}
\begin{feynartspicture}(100,100)(1,1)

\FADiagram{}
\FAProp(0.,15.)(5.5,10.)(0.,){/Straight}{1}
\FALabel(2.18736,11.8331)[tr]{$q_i$}
\FAProp(0.,5.)(5.5,10.)(0.,){/Straight}{-1}
\FALabel(3.31264,6.83309)[tl]{$\bar q_i$}
\FAProp(20.,17.)(11.5,10.)(0.,){/ScalarDash}{0}
\FALabel(15.4036,14.0235)[br]{$\PH$}
\FAProp(20.,10.)(15.5,6.5)(0.,){/Straight}{-1}
\FALabel(17.2784,8.9935)[br]{$\nu_\Pl$}
\FAProp(20.,3.)(15.5,6.5)(0.,){/Straight}{1}
\FALabel(18.2216,5.4935)[bl]{$\bar\nu_\Pl$}
\FAProp(5.5,10.)(11.5,10.)(0.,){/Sine}{0}
\FALabel(8.5,11.07)[b]{$\PZ$}
\FAProp(11.5,10.)(15.5,6.5)(0.,){/Sine}{0}
\FALabel(12.9593,7.56351)[tr]{$\PZ$}
\FAVert(5.5,10.){0}
\FAVert(11.5,10.){0}
\FAVert(15.5,6.5){0}

\end{feynartspicture}

\vspace*{-2.5em}
\end{center}
\mycaption{\label{fi:borndiag} Feynman diagrams for the LO processes \refeq{eq:proc1}--\refeq{eq:proc4}.}
\efi
In detail, the following partonic processes are considered,
\begin{eqnarray}
\label{eq:proc1}
u_i \, \, \bar d_j \,\to & \PH\PW^+ & \to\, \PH \nu_l l^+ \, ,\\
\label{eq:proc2}
d_j \, \, \bar u_i \,\to & \PH\PW^- & \to\, \PH l^- \bar\nu_l\, ,\\
\label{eq:proc3}
q_i \, \, \bar q_i \,\to & \PH\PZ &   \to\, \PH l^- l^+ \, ,\\
\label{eq:proc4}
q_i \, \, \bar q_i \,\to & \PH\PZ &   \to\, \PH \nu_l \bar\nu_l \, ,
\end{eqnarray}
where $q_i$ denotes any light quark and $u_i$, $d_i$ the up- and
down-type quarks of the $i$th generation.  The intermediate W/Z-boson
resonances are described by complex W/Z-boson masses $ \mu_V$ via the
replacement
\begin{equation}
\MV^2 \to \mu_{\PV}^2 = \MV^2 -\ri \MV \GV, \quad \PV=\PW,\PZ
\end{equation}
in the \PV~propagator as dictated by the complex-mass scheme (see
below). Hence, all our results correspond to a fixed-width description
of the Breit--Wigner resonance. Moreover, all related quantities,
in particular the weak mixing angle, are formulated in terms of
the complex mass parameters. 

The final-state leptons are treated as massless unless their small 
masses are used to regularize a collinear divergence.
Concerning bremsstrahlung, we support the possibility that collinear
photons may be completely separated from an outgoing charged
lepton, because this situation is relevant for muons. 
More details on the treatment of such non-collinear-safe observables
are described below.

The light quarks are considered massless as well, in line with the
parton-model requirements. This means that the quark mixing matrix
only appears as global weight factor $|V_{ij}|^2$ in the processes
\refeq{eq:proc1} and \refeq{eq:proc2}. Mixing with the third
generation is neglected throughout.  Owing to the smallness of the
$\Pb\bar\Pb$ contribution to
$\PZ\PH$ production, which amounts to approximately $1(3)\%$
at the 7(14)~TeV LHC for an inclusive analysis, $0.5(1)\%$ 
for a boosted analysis, and is negligible at the Tevatron,
the $\Pb\bar\Pb$ annihilation channels are treated in LO only.

To define the
electromagnetic coupling constant $\alpha$, we use the $\GF$ scheme,
\ie we derive $\alpha$ from the Fermi constant according to
\beq
 \alpha_{\GF} = \frac{\sqrt{2}\GF\MW^2}{\pi}\left(1-\frac{\MW^2}{\MZ^2}\right).
\label{eq:alpgf}
\eeq
In this scheme, the weak corrections to muon decay, $\De r$, are
included in the charge renormalization constant (see \eg
\citere{Ciccolini:2003jy}).  As a consequence, the EW corrections are
independent of logarithms of the light-quark masses. Moreover, this
definition effectively resums the contributions associated with the
running of $\al$ from zero to the weak scale and absorbs some leading
universal corrections $\propto\GF\Mt^2$ from the $\rho$~parameter into
the LO amplitude.
For corrections due to collinear final-state radiation it would be
more appropriate to use $\alpha(0)$ defined in the Thomson limit to
describe the corresponding coupling. On the other hand, using
$\alpha_{\GF}$ everywhere is best suited to describe the 
larger genuine weak corrections which are present in the Higgs-strahlung
processes. Thus,
the optimal choice cannot be achieved in one particular input scheme,
and the calculation could be refined beyond NLO.
Among other things, higher-order effects from
multi-photon emission should also be included at this level of
precision which is beyond the scope of this work. 
However, this kind of sophistication will not be needed
in practice, since the expected statistics does not allow for
this level of accuracy in the experimental analysis either.

The major part of the calculation could be carried over from
the calculation of NLO QCD and EW corrections to Higgs + 2 jet production
as described in \citere{Ciccolini:2007jr}.
In detail, we 
had to discard the vector-boson fusion contribution
($t/u$ channels) and to keep only the $s$-channel part from 
\citere{Ciccolini:2007jr} adjusting the quantum numbers of the final-state
fermions, which are leptons for the Higgs-strahlung processes under consideration. 

All results are checked against a completely new and independent
second calculation based on {\sc FeynArts} 3.2 \cite{Hahn:2000kx} and
{\sc FormCalc} 3.1 \cite{Hahn:1998yk}. The translation of the
amplitudes into the Weyl--van-der-Waerden formalism as presented in
\citere{Dittmaier:1998nn} is performed with the program {\sc
  Pole}~\cite{Accomando:2005ra}. {\sc Pole} also provides an interface
to the multi-channel phase-space integrator {\sc
  Lusifer}~\cite{Dittmaier:2002ap} which has been extended to use {\sc
  Vegas} in order to optimize each phase-space mapping.  The results
of the two calculations are in mutual agreement.

\subsection{Virtual corrections}
\label{se:virt} 

The actual loop calculation, which was also the basis for the
loop calculation~\cite{Ciccolini:2007jr} entering {\sc Hawk},
goes back to the calculation~\cite{Bredenstein:2006rh}
of NLO QCD and EW corrections to the four-body Higgs decays
$\PH\to\PW\PW/\PZ\PZ\to4\,$fermions, where the same
amplitudes appear in a crossed variant.
We make use of those results, which were obtained in the 
traditional Feynman-diagrammatic approach and checked
against the independent second calculation.

The virtual corrections modify the partonic processes that are already
present at LO; there are about 
200 one-loop diagrams per tree diagram
depending on the channel.
At NLO these corrections are induced by
self-energy, vertex, box (4-point), and pentagon (5-point) diagrams,
shown schematically in \reffi{fi:diagVF}.
\bfi
\begin{center}
Self-energy diagrams: \hfill \mbox{} \\[-1.5em]
\begin{feynartspicture}(100,100)(1,1)

\FADiagram{}
\FAProp(0.,15.)(5.5,10.)(0.,){/Straight}{1}
\FALabel(2.18736,11.8331)[tr]{$q_i$}
\FAProp(0.,5.)(5.5,10.)(0.,){/Straight}{-1}
\FALabel(3.31264,6.83309)[tl]{$\bar q_j$}
\FAProp(20.,17.)(11.5,10.)(0.,){/ScalarDash}{0}
\FALabel(15.4036,14.0235)[br]{$\PH$}
\FAProp(20.,10.)(15.5,6.5)(0.,){/Straight}{-1}
\FALabel(17.2784,8.9935)[br]{$\Pl$}
\FAProp(20.,3.)(15.5,6.5)(0.,){/Straight}{1}
\FALabel(15.2216,2.0)[bl]{$\bar \Pl'$}
\FAProp(5.5,10.)(11.5,10.)(0.,){/Sine}{0}
\FALabel(8.5,11.57)[b]{${V}$}
\FAProp(11.5,10.)(15.5,6.5)(0.,){/Sine}{0}
\FALabel(12.9593,6.56351)[tr]{${V}$}
\FAVert(5.5,10.){0}
\FAVert(8.5,10.){-1}
\FAVert(11.5,10.){0}
\FAVert(15.5,6.5){0}

\end{feynartspicture}
\hspace*{0.5em}
\begin{feynartspicture}(100,100)(1,1)

\FADiagram{}
\FAProp(0.,15.)(5.5,10.)(0.,){/Straight}{1}
\FALabel(2.18736,11.8331)[tr]{$q_i$}
\FAProp(0.,5.)(5.5,10.)(0.,){/Straight}{-1}
\FALabel(3.31264,6.83309)[tl]{$\bar q_j$}
\FAProp(20.,17.)(11.5,10.)(0.,){/ScalarDash}{0}
\FALabel(15.4036,14.0235)[br]{$\PH$}
\FAProp(20.,10.)(15.5,6.5)(0.,){/Straight}{-1}
\FALabel(17.2784,8.9935)[br]{$\Pl$}
\FAProp(20.,3.)(15.5,6.5)(0.,){/Straight}{1}
\FALabel(15.2216,2.0)[bl]{$\bar \Pl'$}
\FAProp(5.5,10.)(11.5,10.)(0.,){/Sine}{0}
\FALabel(8.5,11.57)[b]{${V}$}
\FAProp(11.5,10.)(15.5,6.5)(0.,){/Sine}{0}
\FALabel(12.9593,6.56351)[tr]{${V}$}
\FAVert(5.5,10.){0}
\FAVert(11.5,10.){0}
\FAVert(13.5,8.25){-1}
\FAVert(15.5,6.5){0}

\end{feynartspicture}

\mbox{} \\[-2.5em]

Vertex diagrams: \hfill \mbox{} \\[-1.5em]
\begin{feynartspicture}(100,100)(1,1)

\FADiagram{}
\FAProp(0.,15.)(5.5,10.)(0.,){/Straight}{1}
\FALabel(2.18736,11.8331)[tr]{$q_i$}
\FAProp(0.,5.)(5.5,10.)(0.,){/Straight}{-1}
\FALabel(3.31264,6.83309)[tl]{$\bar q_j$}
\FAProp(20.,17.)(11.5,10.)(0.,){/ScalarDash}{0}
\FALabel(15.4036,14.0235)[br]{$\PH$}
\FAProp(20.,10.)(15.5,6.5)(0.,){/Straight}{-1}
\FALabel(17.2784,8.9935)[br]{$\Pl$}
\FAProp(20.,3.)(15.5,6.5)(0.,){/Straight}{1}
\FALabel(15.2216,2.0)[bl]{$\bar \Pl'$}
\FAProp(5.5,10.)(11.5,10.)(0.,){/Sine}{0}
\FALabel(8.5,11.57)[b]{${V}$}
\FAProp(11.5,10.)(15.5,6.5)(0.,){/Sine}{0}
\FALabel(12.9593,6.56351)[tr]{${V}$}
\FAVert(5.5,10.){-1}
\FAVert(11.5,10.){0}
\FAVert(15.5,6.5){0}

\end{feynartspicture}
\hspace*{0.5em}
\begin{feynartspicture}(100,100)(1,1)

\FADiagram{}
\FAProp(0.,15.)(5.5,10.)(0.,){/Straight}{1}
\FALabel(2.18736,11.8331)[tr]{$q_i$}
\FAProp(0.,5.)(5.5,10.)(0.,){/Straight}{-1}
\FALabel(3.31264,6.83309)[tl]{$\bar q_j$}
\FAProp(20.,17.)(11.5,10.)(0.,){/ScalarDash}{0}
\FALabel(15.4036,14.0235)[br]{$\PH$}
\FAProp(20.,10.)(15.5,6.5)(0.,){/Straight}{-1}
\FALabel(17.2784,8.9935)[br]{$\Pl$}
\FAProp(20.,3.)(15.5,6.5)(0.,){/Straight}{1}
\FALabel(15.2216,2.0)[bl]{$\bar \Pl'$}
\FAProp(5.5,10.)(11.5,10.)(0.,){/Sine}{0}
\FALabel(8.5,11.57)[b]{${V}$}
\FAProp(11.5,10.)(15.5,6.5)(0.,){/Sine}{0}
\FALabel(12.9593,6.56351)[tr]{${V}$}
\FAVert(5.5,10.){0}
\FAVert(11.5,10.){-1}
\FAVert(15.5,6.5){0}

\end{feynartspicture}
\hspace*{0.5em}
\begin{feynartspicture}(100,100)(1,1)

\FADiagram{}
\FAProp(0.,15.)(5.5,10.)(0.,){/Straight}{1}
\FALabel(2.18736,11.8331)[tr]{$q_i$}
\FAProp(0.,5.)(5.5,10.)(0.,){/Straight}{-1}
\FALabel(3.31264,6.83309)[tl]{$\bar q_j$}
\FAProp(20.,17.)(11.5,10.)(0.,){/ScalarDash}{0}
\FALabel(15.4036,14.0235)[br]{$\PH$}
\FAProp(20.,10.)(15.5,6.5)(0.,){/Straight}{-1}
\FALabel(17.2784,8.9935)[br]{$\Pl$}
\FAProp(20.,3.)(15.5,6.5)(0.,){/Straight}{1}
\FALabel(15.2216,2.0)[bl]{$\bar \Pl'$}
\FAProp(5.5,10.)(11.5,10.)(0.,){/Sine}{0}
\FALabel(8.5,11.57)[b]{${V}$}
\FAProp(11.5,10.)(15.5,6.5)(0.,){/Sine}{0}
\FALabel(12.9593,6.56351)[tr]{${V}$}
\FAVert(5.5,10.){0}
\FAVert(11.5,10.){0}
\FAVert(15.5,6.5){-1}

\end{feynartspicture}

\hfill \mbox{} \\[-4em]
\begin{feynartspicture}(100,100)(1,1)

\FADiagram{}
\FAProp(0.,15.)(2.75,12.5)(0.,){/Straight}{1}
\FAProp(2.75,12.5)(5.5,10.)(0.,){/Straight}{1}
\FALabel(.5,13.5)[tr]{$q_i$}
\FAProp(0.,5.)(5.5,10.)(0.,){/Straight}{-1}
\FALabel(3.31264,6.83309)[tl]{$\bar q_j$}
\FAProp(7.25,15.5)(2.25,12.5)(0.,){/ScalarDash}{0}
\FALabel(9.4036,14.3235)[br]{$\PH$}
\FAProp(17.,15.)(11.5,10)(0.,){/Straight}{-1}
\FALabel(13.5,13.)[br]{$\Pl$}
\FAProp(17.,5.)(11.5,10)(0.,){/Straight}{1}
\FAProp(5.5,10.)(11.5,10.)(0.,){/Sine}{0}
\FALabel(8.5,11.25)[b]{${V}$}
\FALabel(13.5,6.56351)[tr]{$\bar \Pl'$}
\FAVert(5.5,10.){0}
\FAVert(11.5,10.){0}
\FAVert(2.75,12.5){-1}

\end{feynartspicture}
\hspace*{-0.5em}
\begin{feynartspicture}(100,100)(1,1)

\FADiagram{}
\FAProp(0.,15.)(5.5,10.)(0.,){/Straight}{1}
\FALabel(2.18736,11.8331)[tr]{$q_i$}
\FAProp(0.,5.)(2.75,7.5)(0.,){/Straight}{-1}
\FAProp(2.75,7.5)(5.5,10.)(0.,){/Straight}{-1}
\FALabel(0.5,5.)[tl]{$\bar q_j$}
\FAProp(2.75,7.5)(7.75,4.5)(0.,){/ScalarDash}{0}
\FALabel(9.5,5.5)[br]{$\PH$}
\FAProp(17.,15.)(11.5,10)(0.,){/Straight}{-1}
\FALabel(13.5,13.)[br]{$\Pl$}
\FAProp(17.,5.)(11.5,10)(0.,){/Straight}{1}
\FAProp(5.5,10.)(11.5,10.)(0.,){/Sine}{0}
\FALabel(8.5,11.25)[b]{${V}$}
\FALabel(13.5,6.56351)[tr]{$\bar \Pl'$}
\FAVert(5.5,10.){0}
\FAVert(11.5,10.){0}
\FAVert(2.75,7.5){-1}

\end{feynartspicture}
\hspace*{-0.5em}
\begin{feynartspicture}(100,100)(1,1)

\FADiagram{}
\FAProp(0.,15.)(5.5,10.)(0.,){/Straight}{1}
\FALabel(2.18736,11.8331)[tr]{$q_i$}
\FAProp(0.,5.)(5.5,10.)(0.,){/Straight}{-1}
\FALabel(3.31264,6.83309)[tl]{$\bar q_j$}
\FAProp(14.25,12.5)(17,10)(0.,){/ScalarDash}{0}
\FALabel(18,8)[br]{$\PH$}
\FAProp(14.25,12.5)(11.5,10)(0.,){/Straight}{-1}
\FAProp(17.,15.)(14.25,12.5)(0.,){/Straight}{-1}
\FALabel(16.,15.)[br]{$\Pl$}
\FAProp(17.,5.)(11.5,10)(0.,){/Straight}{1}
\FAProp(5.5,10.)(11.5,10.)(0.,){/Sine}{0}
\FALabel(8.5,11.25)[b]{${V}$}
\FALabel(13.5,6.56351)[tr]{$\bar \Pl'$}
\FAVert(5.5,10.){0}
\FAVert(11.5,10.){0}
\FAVert(14.25,12.5){-1}

\end{feynartspicture}
\hspace*{-0.5em}
\begin{feynartspicture}(100,100)(1,1)

\FADiagram{}
\FAProp(0.,15.)(5.5,10.)(0.,){/Straight}{1}
\FALabel(2.18736,11.8331)[tr]{$q_i$}
\FAProp(0.,5.)(5.5,10.)(0.,){/Straight}{-1}
\FALabel(3.31264,6.83309)[tl]{$\bar q_j$}
\FAProp(14.25,7.5)(17,10)(0.,){/ScalarDash}{0}
\FALabel(18,10.25)[br]{$\PH$}
\FAProp(17.,15.)(11.5,10)(0.,){/Straight}{-1}
\FALabel(13.5,13.)[br]{$\Pl$}
\FAProp(14.25,7.5)(11.5,10)(0.,){/Straight}{1}
\FAProp(17.,5.)(14.25,7.5)(0.,){/Straight}{1}
\FAProp(5.5,10.)(11.5,10.)(0.,){/Sine}{0}
\FALabel(8.5,11.25)[b]{${V}$}
\FALabel(15.75,5)[tr]{$\bar \Pl'$}
\FAVert(5.5,10.){0}
\FAVert(11.5,10.){0}
\FAVert(14.25,7.5){-1}

\end{feynartspicture}

\mbox{} \\[-2.5em]

Box and pentagon diagrams: \hfill \mbox{} \\[-1em]
\begin{feynartspicture}(100,100)(1,1)

\FADiagram{}
\FAProp(0.,15.)(5.5,10.)(0.,){/Straight}{1}
\FALabel(2.18736,11.8331)[tr]{$q_i$}
\FAProp(0.,5.)(5.5,10.)(0.,){/Straight}{-1}
\FALabel(3.31264,6.83309)[tl]{$\bar q_j$}
\FAProp(20.,17.)(12,10)(0.,){/ScalarDash}{0}
\FALabel(17.2784,15.9935)[br]{$\PH$}
\FAProp(20.,10.)(12,10)(0.,){/Straight}{-1}
\FALabel(18.2216,10.4935)[bl]{$\Pl$}
\FAProp(20.,3.)(12.,10.)(0.,){/Straight}{1}
\FALabel(15.4593,5.81351)[tr]{$\bar \Pl'$}
\FAProp(5.5,10.)(12.,10.)(0.,){/Sine}{0}
\FALabel(8.75,8.93)[t]{${V}$}
\FAVert(5.5,10.){0}
\FAVert(12.,10.){-1}

\end{feynartspicture}
\hspace*{0.5em}
\begin{feynartspicture}(100,100)(1,1)

\FADiagram{}
\FAProp(0.,15.)(10,10.)(0.,){/Straight}{1}
\FALabel(4.18736,11.8331)[tr]{$q_i$}
\FAProp(0.,5.)(10,10.)(0.,){/Straight}{-1}
\FALabel(3.31264,5.83309)[tl]{$\bar q_j$}
\FAProp(20.,17.)(15.5,13.5)(0.,){/Straight}{1}
\FALabel(17.2784,15.9935)[br]{$\Pl$}
\FAProp(20.,10.)(15.5,13.5)(0.,){/Straight}{-1}
\FALabel(18.2216,7.4935)[bl]{$\bar \Pl'$}
\FAProp(20.,3.)(10,10.)(0.,){/ScalarDash}{0}
\FALabel(13,5)[tl]{$\PH$}
\FAProp(15.5,13.5)(10,10.)(0.,){/Sine}{0}
\FALabel(13.134,12.366)[br]{${V}$}
\FAVert(15.5,13.5){0}
\FAVert(10,10.){-1}

\end{feynartspicture}
\hspace*{0.5em}
\begin{feynartspicture}(100,100)(1,1)

\FADiagram{}
\FAProp(0.,15.)(10,10.)(0.,){/Straight}{1}
\FALabel(4.18736,11.8331)[tr]{$q_i$}
\FAProp(0.,5.)(10,10.)(0.,){/Straight}{-1}
\FALabel(3.31264,5.83309)[tl]{$\bar q_j$}
\FAProp(20,3)(10,10)(0,){/Straight}{1}
\FALabel(13,6)[tl]{$\bar \Pl'$}
\FAProp(20.,17.)(10,10)(0.,){/ScalarDash}{0}
\FALabel(17.4593,15.9365)[br]{$\PH$}
\FAProp(20.,10.)(10,10)(0.,){/Straight}{-1}
\FALabel(17.5407,10.4365)[bl]{$\Pl$}
\FAVert(10.,10.){-1}

\end{feynartspicture}

\vspace*{-1.6em}
\end{center}
\mycaption{\label{fi:diagVF} Different classes of loop diagrams
with up to 5 external legs corresponding to the virtual corrections to
the LO processes \refeq{eq:proc1}--\refeq{eq:proc4}. Note that there are
loop-induced vertex diagrams for the coupling of the Higgs boson to the external 
massless fermions which are absent at tree level.}
\efi
For the implementation of the finite W/Z widths the complex-mass
scheme is employed, which was introduced in \citere{Denner:1999gp} for
LO calculations and generalized to the one-loop level in
\citere{Denner:2005fg}. In this approach the W- and Z-boson masses are
consistently considered as complex quantities, defined as the
locations of the propagator poles in the complex plane.  This leads to
complex couplings and, in particular, a complex weak mixing angle.
The scheme fully respects all relations that follow from gauge
invariance.

The underlying calculation~\cite{Bredenstein:2006rh} of the
$\PH\to4f$ amplitudes was performed both in
the conventional 't~Hooft--Feynman gauge and in the background-field
formalism using the conventions of \citeres{Denner:1991kt} and
\cite{Denner:1994xt}, respectively.
The amplitudes were generated with {\sc FeynArts 1.0} \cite{Kublbeck:1990xc}
and further manipulated with in-house programs written in 
{\sl Mathematica}, which automatically create building blocks in
{\sl Fortran}. The amplitudes are expressed in terms
of standard matrix elements, which contain the fermionic spinor chains,
and coefficients, which contain the tensor integrals, couplings, etc.
The tensor integrals 
are recursively reduced to scalar master integrals at the numerical
level. 
Tensor and scalar 5-point
functions are directly expressed in terms of 4-point integrals
\cite{Denner:2002ii,Denner:2005nn}.  Tensor 4-point and 3-point
integrals are reduced to scalar integrals with the Passarino--Veltman
algorithm \cite{Passarino:1979jh} as long as no small Gram determinant
appears in the reduction. 
If small Gram determinants occur, we expand
the tensor coefficients about the limit of vanishing Gram determinants
and possibly other kinematical determinants, as described in
\citere{Denner:2005nn} in detail.
For the evaluation of the scalar integrals with complex masses
we use two independent in-house libraries based on the general results 
listed in \citere{Denner:2010tr}.

\subsection{Real corrections}
\label{se:real} 

The EW real corrections to the partonic processes
\refeq{eq:proc1}--\refeq{eq:proc4} consist of photon
bremsstrahlung,
\begin{eqnarray}
\label{eq:proc5}
u_i \, \, \bar d_j \,\to & \PH\PW^+ & \to\, \PH \nu_l l^+ \ga\, ,\\
\label{eq:proc6}
d_j \, \, \bar u_i \,\to & \PH\PW^- & \to\, \PH l^- \bar\nu_l\ga\, ,\\
\label{eq:proc7}
q_i \, \, \bar q_i \,\to & \PH\PZ &   \to\, \PH l^- l^+ \ga\, ,\\
\label{eq:proc8}
q_i \, \, \bar q_i \,\to & \PH\PZ &   \to\, \PH \nu_l \bar\nu_l \ga\, ,
\end{eqnarray}
with Feynman diagrams shown exemplarily for the process
\refeq{eq:proc5} in \reffi{fi:realdiags}, and of photon-induced
processes, which involve the same amplitudes as the bremsstrahlung
processes, but with the photon crossed into the initial state in all
possible ways.
\bfi[t]
\begin{center}
\begin{feynartspicture}(100,100)(1,1)

\FADiagram{}
\FAProp(0.,15.)(2.75,12.5)(0.,){/Straight}{1}
\FAProp(2.75,12.5)(5.5,10.)(0.,){/Straight}{1}
\FALabel(2.18736,11.8331)[tr]{$u_i$}
\FAProp(0.,5.)(5.5,10.)(0.,){/Straight}{-1}
\FALabel(3.31264,6.83309)[tl]{$\bar d_j$}
\FAProp(20.,17.)(11.5,10.)(0.,){/ScalarDash}{0}
\FALabel(15.4036,14.0235)[br]{$\PH$}
\FAProp(20.,10.)(15.5,6.5)(0.,){/Straight}{-1}
\FALabel(17.2784,8.9935)[br]{$\nu_\Pl$}
\FAProp(20.,3.)(15.5,6.5)(0.,){/Straight}{1}
\FALabel(18.2216,5.4935)[bl]{$\Pl^+$}
\FAProp(5.5,10.)(11.5,10.)(0.,){/Sine}{0}
\FALabel(8.5,11.07)[b]{$\PW$}
\FAProp(11.5,10.)(15.5,6.5)(0.,){/Sine}{0}
\FALabel(12.9593,7.56351)[tr]{$\PW$}
\FAVert(5.5,10.){0}
\FAVert(11.5,10.){0}
\FAVert(15.5,6.5){0}
\FAProp(2.75,12.5)(8.25,17.)(0.,){/Sine}{0}
\FAVert(2.75,12.5){0}
\FALabel(5.,16.5)[tr]{$\gamma$}

\end{feynartspicture}
\hspace*{.5em}
\begin{feynartspicture}(100,100)(1,1)

\FADiagram{}
\FAProp(0.,15.)(5.5,10)(0.,){/Straight}{1}
\FALabel(4.,15.5)[tr]{$u_i$}
\FAProp(0.,5.)(2.75,7.5)(0.,){/Straight}{-1}
\FAProp(2.75,7.5)(5.5,10.)(0.,){/Straight}{-1}
\FALabel(0.25,10.5)[tl]{$\bar d_j$}
\FAProp(20.,17.)(11.5,10.)(0.,){/ScalarDash}{0}
\FALabel(15.4036,14.0235)[br]{$\PH$}
\FAProp(20.,10.)(15.5,6.5)(0.,){/Straight}{-1}
\FALabel(17.2784,8.9935)[br]{$\nu_\Pl$}
\FAProp(20.,3.)(15.5,6.5)(0.,){/Straight}{1}
\FALabel(18.2216,5.4935)[bl]{$\Pl^+$}
\FAProp(5.5,10.)(11.5,10.)(0.,){/Sine}{0}
\FALabel(8.5,11.07)[b]{$\PW$}
\FAProp(11.5,10.)(15.5,6.5)(0.,){/Sine}{0}
\FALabel(12.9593,7.56351)[tr]{$\PW$}
\FAVert(5.5,10.){0}
\FAVert(11.5,10.){0}
\FAVert(15.5,6.5){0}
\FAProp(2.75,7.5)(8.25,2.)(0.,){/Sine}{0}
\FAVert(2.75,7.5){0}
\FALabel(5.,4.)[tr]{$\gamma$}

\end{feynartspicture}
\hspace*{.5em}
\begin{feynartspicture}(100,100)(1,1)

\FADiagram{}
\FAProp(0.,15.)(5.5,10.)(0.,){/Straight}{1}
\FALabel(2.18736,11.8331)[tr]{$u_i$}
\FAProp(0.,5.)(5.5,10.)(0.,){/Straight}{-1}
\FALabel(3.31264,6.83309)[tl]{$\bar d_j$}
\FAProp(20.,17.)(11.5,10.)(0.,){/ScalarDash}{0}
\FALabel(15.4036,14.0235)[br]{$\PH$}
\FAProp(20.,10.)(15.5,6.5)(0.,){/Straight}{-1}
\FALabel(17.2784,8.9935)[br]{$\nu_\Pl$}
\FAProp(20.,3.)(17.75,4.75)(0.,){/Straight}{1}
\FAProp(17.75,4.75)(15.5,6.5)(0.,){/Straight}{1}
\FALabel(14.5,3.)[bl]{$\Pl^+$}
\FAProp(5.5,10.)(11.5,10.)(0.,){/Sine}{0}
\FALabel(8.5,11.07)[b]{$\PW$}
\FAProp(11.5,10.)(15.5,6.5)(0.,){/Sine}{0}
\FALabel(12.9593,7.56351)[tr]{$\PW$}
\FAVert(5.5,10.){0}
\FAVert(11.5,10.){0}
\FAVert(15.5,6.5){0}
\FAProp(17.75,4.75)(22.,6.5)(0.,){/Sine}{0}
\FAVert(17.75,4.75){0}
\FALabel(22.,5.)[tr]{$\gamma$}

\end{feynartspicture}
\hspace*{.5em}
\begin{feynartspicture}(100,100)(1,1)

\FADiagram{}
\FAProp(0.,15.)(5.5,10.)(0.,){/Straight}{1}
\FALabel(2.18736,11.8331)[tr]{$u_i$}
\FAProp(0.,5.)(5.5,10.)(0.,){/Straight}{-1}
\FALabel(3.31264,6.83309)[tl]{$\bar d_j$}
\FAProp(20.,17.)(11.5,10.)(0.,){/ScalarDash}{0}
\FALabel(15.4036,14.0235)[br]{$\PH$}
\FAProp(20.,10.)(15.5,6.5)(0.,){/Straight}{-1}
\FALabel(17.2784,8.9935)[br]{$\nu_\Pl$}
\FAProp(20.,3.)(15.5,6.5)(0.,){/Straight}{1}
\FALabel(18.2216,5.4935)[bl]{$\Pl^+$}
\FAProp(5.5,10.)(11.5,10.)(0.,){/Sine}{0}
\FALabel(7.,7.25)[b]{$\PW$}
\FALabel(10.,7.25)[b]{$\PW$}
\FAProp(11.5,10.)(15.5,6.5)(0.,){/Sine}{0}
\FALabel(14.25,7.)[tr]{$\PW$}
\FAVert(5.5,10.){0}
\FAVert(11.5,10.){0}
\FAVert(15.5,6.5){0}
\FAProp(8.5,10.)(13,17.)(0.,){/Sine}{0}
\FAVert(8.5,10.){0}
\FALabel(11.,16.5)[tr]{$\gamma$}

\end{feynartspicture}

\vspace*{-2.em}

\begin{feynartspicture}(100,100)(1,1)

\FADiagram{}
\FAProp(0.,15.)(5.5,10.)(0.,){/Straight}{1}
\FALabel(2.18736,11.8331)[tr]{$u_i$}
\FAProp(0.,5.)(5.5,10.)(0.,){/Straight}{-1}
\FALabel(3.31264,6.83309)[tl]{$\bar d_j$}
\FAProp(20.,17.)(11.5,10.)(0.,){/ScalarDash}{0}
\FALabel(15.4036,14.0235)[br]{$\PH$}
\FAProp(20.,10.)(15.5,6.5)(0.,){/Straight}{-1}
\FALabel(19.2784,9.9935)[br]{$\nu_\Pl$}
\FAProp(20.,3.)(15.5,6.5)(0.,){/Straight}{1}
\FALabel(18.2216,5.4935)[bl]{$\Pl^+$}
\FAProp(5.5,10.)(11.5,10.)(0.,){/Sine}{0}
\FALabel(8.5,11.07)[b]{$\PW$}
\FAProp(11.5,10.)(15.5,6.5)(0.,){/Sine}{0}
\FALabel(12.75,8)[tr]{$\PW$}
\FAVert(5.5,10.){0}
\FAVert(11.5,10.){0}
\FAVert(15.5,6.5){0}
\FAProp(13.5,8.25)(14,3.)(0.,){/Sine}{0}
\FAVert(13.5,8.25){0}
\FALabel(13.,5)[tr]{$\gamma$}
\FALabel(16.5,9.5)[tr]{$\PW$}

\end{feynartspicture}
\hspace*{.5em}
\begin{feynartspicture}(100,100)(1,1)

\FADiagram{}
\FAProp(0.,15.)(5.5,10.)(0.,){/Straight}{1}
\FALabel(2.18736,11.8331)[tr]{$u_i$}
\FAProp(0.,5.)(5.5,10.)(0.,){/Straight}{-1}
\FALabel(3.31264,6.83309)[tl]{$\bar d_j$}
\FAProp(20.,17.)(11.5,10.)(0.,){/ScalarDash}{0}
\FALabel(15.4036,14.0235)[br]{$\PH$}
\FAProp(20.,10.)(15.5,6.5)(0.,){/Straight}{-1}
\FALabel(17.2784,8.9935)[br]{$\nu_\Pl$}
\FAProp(20.,3.)(15.5,6.5)(0.,){/Straight}{1}
\FALabel(18.2216,5.4935)[bl]{$\Pl^+$}
\FAProp(5.5,10.)(8.5,10.)(0.,){/Sine}{0}
\FAProp(8.5,10.)(11.5,10.)(0.,){/ScalarDash}{0}
\FALabel(7.,7.25)[b]{$\PW$}
\FALabel(10.,7.25)[b]{$\phi$}
\FAProp(11.5,10.)(15.5,6.5)(0.,){/Sine}{0}
\FALabel(13.75,6.75)[tr]{$\PW$}
\FAVert(5.5,10.){0}
\FAVert(11.5,10.){0}
\FAVert(15.5,6.5){0}
\FAProp(8.5,10.)(13,17.)(0.,){/Sine}{0}
\FAVert(8.5,10.){0}
\FALabel(11.,16.5)[tr]{$\gamma$}

\end{feynartspicture}
\hspace*{.5em}
\begin{feynartspicture}(100,100)(1,1)

\FADiagram{}
\FAProp(0.,15.)(5.5,10.)(0.,){/Straight}{1}
\FALabel(2.18736,11.8331)[tr]{$u_i$}
\FAProp(0.,5.)(5.5,10.)(0.,){/Straight}{-1}
\FALabel(3.31264,6.83309)[tl]{$\bar d_j$}
\FAProp(20.,17.)(11.5,10.)(0.,){/ScalarDash}{0}
\FALabel(15.4036,14.0235)[br]{$\PH$}
\FAProp(20.,10.)(15.5,6.5)(0.,){/Straight}{-1}
\FALabel(19.2784,9.9935)[br]{$\nu_\Pl$}
\FAProp(20.,3.)(15.5,6.5)(0.,){/Straight}{1}
\FALabel(18.2216,5.4935)[bl]{$\Pl^+$}
\FAProp(5.5,10.)(11.5,10.)(0.,){/Sine}{0}
\FALabel(8.5,11.07)[b]{$\PW$}
\FAProp(11.5,10.)(13.5,8.25)(0.,){/ScalarDash}{0}
\FAProp(13.5,8.25)(15.5,6.5)(0.,){/Sine}{0}
\FALabel(12.25,8.75)[tr]{$\phi$}
\FAVert(5.5,10.){0}
\FAVert(11.5,10.){0}
\FAVert(15.5,6.5){0}
\FAProp(13.5,8.25)(14,3.)(0.,){/Sine}{0}
\FAVert(13.5,8.25){0}
\FALabel(13.,5)[tr]{$\gamma$}
\FALabel(16.5,9.5)[tr]{$\PW$}

\end{feynartspicture}

\vspace*{-2.5em}
\end{center}
\mycaption{\label{fi:realdiags}
Feynman diagrams for real photonic bremsstrahlung corrections 
to the LO process \refeq{eq:proc1}. A photon cannot only be emitted 
from the external charged fermions, but also from the intermediate \PW\
bosons. In Feynman gauge, there are also diagrams involving the 
charged Goldstone boson $\phi$.}
\efi
The technical implementation of the real corrections again
basically follows along the same lines as in
the NLO calculation within {\sc Hawk}
for 
Higgs + 2 jet 
production~\cite{Ciccolini:2007jr, hawk},
employing dipole subtraction~\cite{Catani:1996vz,Dittmaier:1999mb,Dittmaier:2008md}.

Due to the emission of soft photons the real corrections include soft
singularities which are cancelled by the virtual corrections
independently of the details of the event selection or recombination
procedure.  The treatment of left-over collinear singularities related
to splittings involving photons and fermions requires particular care:
\begin{itemize}
\item {\it Photonic initial-state radiation}\\
  The collinear singularities from photonic initial-state radiation
  off quarks are absorbed by a redefinition of the PDFs similar to the
  standard factorization procedure in QCD, but employing the DIS
  scheme.  At present, the PDF set MRSTQED2004~\cite{Martin:2004dh} is
  the only one that takes into account ${\cal O}(\alpha)$ corrections.
  However, applications with this PDF set and older estimate of ${\cal
    O}(\alpha)$ effects on PDFs show that the impact of these PDF
  corrections is at most at the per-cent
  level~\cite{Spiesberger:1994dm}.  For the time being, it is, thus,
  advisable to use an up-to-date PDF set that ignores ${\cal
    O}(\alpha)$ corrections, but includes recent improvements on the
  QCD side, because the latter have an impact on PDFs much larger than
  1\%.
\item {\it Photon-induced processes, $\gamma\to q\bar q^*/\bar qq^*$ splitting}\\
  In the photon-induced processes, the collinear singularities
  resulting from the splittings $\gamma\to q\bar q^*/\bar qq^*$ are
  removed in the process of PDF redefinitions as well.  Moreover, the
  evaluation of the photon-induced channels requires a photon
  distribution function and its inclusion in the PDF evolution. Using
  the MRSTQED2004 PDF set for this task is certainly legitimate, since
  the contribution from initial-state photons is comparably small.
\item {\it Collinear final-state radiation}\\
Next, we consider collinear final-state radiation off leptons.
If collinear photons and charged
leptons are recombined into a pseudo-particle (mimicking the
start of hadronic or electromagnetic showers) to form IR-safe
observables, all the remaining singularities arising from collinear
photon emission in the final state also cancel against the
corresponding singularities in the virtual corrections. 
For muons in the final state, however, it is experimentally possible
to separate collinear photons from the lepton, i.e.\ to observe
so-called ``bare'' muons. Hence, the resulting cross sections are not
collinear safe (i.e.\ the KLN theorem \cite{Kinoshita:1962ur} does not
apply), and the corresponding collinear singularities show up as
logarithms of the small lepton (muon) mass. The lepton mass cuts off
the collinear divergence in a physically meaningful way.  We employ
the extension \cite{Dittmaier:2008md} of the subtraction formalism
\cite{Catani:1996vz,Dittmaier:1999mb}, which allows for calculating
cross sections for bare leptons, i.e.\ cross sections defined without
any photon recombination. Like in the standard subtraction formalism,
it is sufficient to calculate the real-emission matrix elements for
the partonic processes in the massless-fermion approximation. The main
difference between the subtraction variants of
\citeres{Dittmaier:1999mb} and \cite{Dittmaier:2008md} concerns the
implementation of phase-space cuts. In the standard subtraction
formalism \cite{Dittmaier:1999mb} it is always assumed that the
complete momentum of the lepton and a collinear photon is subject to
cuts (as it would be the case after recombination), while the
generalization \cite{Dittmaier:2008md} allows for non-collinear-safe
cuts that resolve the distribution of the momenta in collinear
photon--lepton configurations. This more general cut procedure in the
non-collinear-safe case has to be carefully implemented both in the
real-emission part and the corresponding subtraction terms, in order
to ensure the numerical cancellation of singularities. Of course, the
treatment of non-collinear-safe cuts leads to a modification of the
re-added subtraction part as well. In the formulation of
\citere{Dittmaier:2008md} this modification assumes the form of an
additional (+)-distribution which contains the surviving mass
singularity. This (+)-distribution integrates to zero for
collinear-safe observables so that the formalism reduces to the
well-known standard subtraction formalism.  For non-collinear-safe
observables the additional logarithms of the lepton mass in the final
result are, thus, isolated analytically.  For a complete and detailed
description of this more general subtraction formalism we refer the
reader to \citere{Dittmaier:2008md}.

\item {\it Photon-induced processes, $\gamma\to l\bar l^*/\bar ll^*$ splitting}\\
  Finally, there are also collinear configurations connected with the
splittings $\gamma\to l\bar l^*/\bar ll^*$ if charged leptons are
allowed to escape into the beam pipe.  For instance, in our numerical
analysis in Section~\ref{se:numres}, we also provide predictions for a
contribution to the $\PH + \nu_\Pl \bar\nu_\Pl$ channels which stems
from $\PW\PH$ production, where the charged lepton is not identified
in the detector. While the EW corrections are conceptually not
different for the quark-induced processes, the photon-induced
processes have to be treated carefully, in particular the contribution
where the incoming photon is splitting into a charged lepton pair,
which leads to an enhancement by a logarithm in the lepton mass.  The
subtraction formalism to extract these logarithms analytically is
available~\cite{Dittmaier:2008md} and proceeds as for the photonic
splitting into quarks.  However, in contrast to the splitting into
quarks the mass logarithm is not absorbed into the definition of the
PDF (there is no lepton PDF), but is part of the result. For the
numerics in \refse{se:numres}, we use the mass of the electron in this
logarithmically enhanced term.
\end{itemize}

\section{Numerical results}
\label{se:numres}

\subsection{Input parameters and setup}
\label{se:SMinput}

For the numerical evaluation we adopt the input parameters
used by the LHC Higgs Cross Section Working Group in
\citere{Dittmaier:2011ti},
\begin{equation}\arraycolsep 2pt
\begin{array}[b]{rclrclrcl}
\GF & = & 1.16637 \times 10^{-5} \GeV^{-2}, \quad&
\alpha_{\mathrm{s}}(\MZ)|_{\NLO} &=& 0.1202 , 
\\
\MW^{\OS} & = & 80.398\GeV, &
\Gamma_\PW^{\OS} & = & 2.0887\GeV, \\
\MZ^{\OS} & = & 91.1876\GeV, &
\Gamma_\PZ^{\OS} & = & 2.4952\GeV, &
M_\PH & = & 120\GeV, \\
m_\Pe & = & 0.510998910\MeV, &
m_\mu &=& 105.658367\MeV,\quad &
m_\Pt & = & 172.5\;\GeV,
\\
|V_{\Pu\Pd}| & = & |V_{\Pc\Ps}| = 0.974, &
|V_{\Pu\Ps}| & = & |V_{\Pc\Pd}| = \sqrt{1 - |V_{\Pc\Ps}|^2}, 
\\[-4ex]
\end{array}
\label{eq:SMpar}
\vspace{2ex}
\end{equation}
which essentially follow \citere{Amsler:2008zzb}. Note that we use the
measured width $\Gamma_\PZ^{\OS}$ for the \PZ~boson, but the
calculated on-shell width $\Gamma_\PW^{\OS}$ for the \PW~boson as
input. The CKM matrix is included via global factors in the partonic
cross sections for the different possible quark flavours.  The small
mixing between the first two and the third generation is neglected.
Within loops the CKM matrix is set to unity, because its effect is
negligible there.

Using the complex-mass scheme~\cite{Denner:2005fg}, we employ a fixed
width in the resonant W- and Z-boson propagators in contrast to the
approach used at LEP and Tevatron
to fit the W~and Z~resonances, where running
widths are taken. Therefore, we have to convert the ``on-shell'' (OS)
values of $M_V^{\OS}$ and $\Ga_V^{\OS}$ ($V=\PW,\PZ$), resulting
from LEP and Tevatron, to the ``pole values'' denoted by $M_V$ and $\Ga_V$. The
relation between the two sets of values is given by~\cite{Bardin:1988xt}
\beq\label{eq:m_ga_pole}
M_V = M_V^{\OS}/
\sqrt{1+(\Ga_V^{\OS}/M_V^{\OS})^2},
\qquad
\Ga_V = \Ga_V^{\OS}/
\sqrt{1+(\Ga_V^{\OS}/M_V^{\OS})^2},
\eeq
leading to
\beqar
\begin{array}[b]{r@{\,}l@{\qquad}r@{\,}l}
\MW &= 80.370\ldots\GeV, & \GW &= 2.0880\ldots\GeV, \\
\MZ &= 91.153\ldots\GeV,& \GZ &= 2.4943\ldots\GeV.
\label{eq:m_ga_pole_num}
\end{array}
\eeqar
We make use of these mass and width parameters in the numerics discussed below,
although the difference between using $M_V$ or $M_V^{\OS}$ would be
hardly visible.

As explained in \refse{se:setup}, we adopt the $\GF$ scheme, where the
electromagnetic coupling $\alpha$ is set to $\alpha_{\GF}$ \refeq{eq:alpgf}.
In this scheme the electric-charge renormalization constant does not contain
logarithms of the light-fermion masses, in contrast to the $\alpha(0)$
scheme, so that the results become practically
independent of the light-quark masses.

The MSTW2008NLO/LO PDF set~\cite{Martin:2009iq} is used throughout
implying the values of $\alpha_{\mathrm{s}}(\MZ)$ stated in
\refeq{eq:SMpar} at NLO.  We use standard running of the strong
coupling constant as provided by the LHAPDF implementation of the PDF
sets~\cite{Whalley:2005nh}.  Only the photon-induced processes are
evaluated with the MRSTQED2004 set of PDFs~\cite{Martin:2004dh}. 
The QCD and QED factorization scales as well as the renormalization
scale are always identified and set to
\begin{equation}
\mu = \MV + \MH \, .
\end{equation}
For a study of the scale and PDF uncertainties in Higgs-strahlung
processes, we refer the reader to the report of the LHC Higgs Cross
Section Working Group~\cite{Dittmaier:2011ti} for total cross sections
and to the upcoming second report~\cite{YR2} on differential
distributions.

In the following, we present results for the Tevatron at a CM energy of
$\sqrt{s}=1.96\TeV$ and for the LHC with $\sqrt{s}=7\TeV$ and $14\TeV$.

\subsection{Phase-space cuts and event selection}
\label{se:cuts}

Since $\PW\PH$ and $\PZ\PH$ production with leptonically decaying W
and Z bosons do not involve outgoing QCD partons in LO, there is at
most one QCD parton in the final state in our fixed-order NLO
predictions, so that no jet algorithm has to be applied.

As explained above, we alternatively apply two versions of
handling photons that can become collinear to outgoing
charged leptons. The first option is to assume a perfect
isolation between charged leptons and photons, an assumption that
is at least approximately fulfilled for (bare) muons.
The second option performs a recombination of photons
and nearly collinear charged leptons and, thus, mimics the
inclusive treatment of electrons within electromagnetic showers in the
detector. Specifically,
a photon $\ga$ and a lepton $l$ are recombined for $R_{\gamma l} < 0.1$,
where $R_{\gamma l} = \sqrt{(y_{l}-y_{\ga})^2+\phi_{l\ga}^2}$
is the usual separation variable in the $y{-}\phi$-plane 
with $y$ denoting the rapidity and
$\phi_{l\ga}$ the angle between $l$ and $\ga$ in 
the plane perpendicular to the beams.
If $l$ and $\ga$ are recombined, we simply add their four-momenta
and treat the resulting object as a quasi-lepton. 
If more than one charged lepton is present in the final state,
the possible recombination is performed with the lepton 
delivering the smaller value of $R_{\gamma l}$.

After applying the recombination procedure we impose the following
cuts on the charged leptons,
\beq
\label{eq:chraged_lepton_cuts}
p_{\mathrm{T},l}>20\GeV, \qquad
|y_l|<2.5,
\eeq
where $p_{\mathrm{T},l}$ is the transverse momentum of
lepton $l$. For channels with at least one neutrino in the final 
state we require a missing transverse momentum 
\beq
\dsl{p}_{\mathrm{T}}>25\GeV \, ,
\eeq
which is equal to the total transverse momentum of the neutrinos in
the events.  While the two-body decay of the Higgs boson is
implemented in {\sc Hawk}, in this work we do not employ
identification cuts on the decay products of the Higgs boson.

For the LHC, we also discuss the impact of the optional additional cuts
\beq
p_{\mathrm{T,H}}>200\GeV, \qquad p_{\mathrm{T,W/Z}}>190\GeV
\eeq
on the transverse momentum of the Higgs and the weak gauge boson,
respectively.  The corresponding selection of events with boosted
Higgs bosons is improving the signal-to-background ratio in the
context of employing the measurement of the jet substructure in $\PH
\to \Pb \bar \Pb$ decays leading to a single fat jet.  The need for
background suppression calls for (almost) identical cuts on the
transverse momentum of the vector bosons and the Higgs
boson~\cite{privcomm}. However, symmetric cuts induce large radiative
corrections in fixed-order calculations in the corresponding
$p_{\mathrm{T}}$ distributions near the cut. Since the Higgs boson and
the vector boson are back-to-back at LO, any initial-state radiation
will either decrease $p_{\mathrm{T},\PH}$ or $p_{\mathrm{T},\PW/\PZ}$,
and the event may not pass the cut anymore. Hence, the differential
cross section near the cut is sensitive to almost collinear and/or
rather soft initial-state radiation. By choosing the above (slightly
asymmetric) cuts this large sensitivity to higher-order corrections
can be removed for the important $p_{\mathrm{T},\PH}$-distribution. Of
course, since the LO distribution for $p_{\mathrm{T},\PW/\PZ}$ is
vanishing for $p_{\mathrm{T},\PW/\PZ}< 200$~GeV due to the
$p_{\mathrm{T},\PH}$ cut, the higher-order corrections to the
$p_{\mathrm{T},\PW/\PZ}$ distributions are still large in this region.

The channel with a Higgs boson and only missing $p_\mathrm{T}$ also
receives a contribution from \PW\PH\ production where the charged
lepton is not identified, which we label $\PH \nu_\Pl /
\overline{\nu}_\Pl$. In our setup, a lepton is not identified if it
fails to pass either the $p_{\mathrm{T},\Pl}$ or the $y_\Pl$ cut
in~(\ref{eq:chraged_lepton_cuts}). Following our generic setup, we
still consider only the neutrino momentum to calculate
$\dsl{p}_{\mathrm{T}}$. Since in this case $\dsl{p}_{\mathrm{T}}$
and $p_{\mathrm{T},\PW}$ are indistinguishable, we also discard the
momentum of the charged lepton when calculating the transverse
momentum of the vector boson, i.e.\ we set $p_{\mathrm{T},\PV} =
\dsl{p}_{\mathrm{T}}$ in this particular case.

\subsection{Results for the LHC and the Tevatron}
\label{se:lhcresults}

In this section, we present numerical results for (differential) cross sections
and various corrections to the Higgs-production channels with one charged lepton 
and missing transverse momentum 
($\PH \Pl^+\nu_\Pl$ and $\PH \Pl^-\overline{\nu}_\Pl$), with two charged
leptons ($\PH \Pl^+\Pl^-$), and with missing momentum only 
in the final state. For the latter, we distinguish the 
\PZ-mediated process
($\PH \nu_\Pl \overline{\nu}_\Pl$) and the 
\PW-mediated ($\PH \nu_\Pl / \overline{\nu}_\Pl$)
process, where the charged lepton is not measured in the detector. Of course, the 
two sources are not distinguishable experimentally. However, 
since the contribution of the two sources varies considerably with the applied 
identification cuts, we state the results separately. For the $\PH \nu_\Pl / \overline{\nu}_\Pl$ 
channel we simply invert the identification cuts (\ref{eq:chraged_lepton_cuts}) for 
the charged lepton 
(see discussion at the end of~\refse{se:cuts}) 
while all other cuts and the recombination procedure is applied
without changes (see Section~\ref{se:real} for subtleties concerning the photon-induced
processes). 
Thus, the resulting cross section may serve as an upper bound on the 
contribution to the channel 
with a Higgs boson and only missing transverse momentum in the
experimental analysis.

Concerning the EW corrections, we distinguish bare muons and final
states with lepton--photon recombination, as introduced in
Section~\ref{se:cuts}.  The corresponding corrections are labelled
$\delta_\mathrm{EW}^{\mathrm{bare}}$ and
$\delta_\mathrm{EW}^{\mathrm{rec}}$, respectively. While the latter is
independent of the specific charged lepton in the final state,
$\delta_\mathrm{EW}^{\mathrm{bare}}$ contains fermion-mass logarithms
and depends on the lepton mass which is taken to be the muon mass. The
contributions due to photon-induced processes are also given in terms
of their relative size with respect to the LO cross section and
are labelled $\delta_\gamma$.

All cross sections are given for a specific leptonic state (e.g.\ 
$\mathrm{e}^+\mathrm{e}^-$, $\mu^+\mu^-$, or $\nu_\mu\bar\nu_\mu$).
While the bare corrections are only applicable for the muon final
state, the results with lepton--photon recombination are equally valid
for electrons and muons. For the process $\PH \nu_\Pl
\overline{\nu}_\Pl$, there is of course no recombination to be
applied, \ie
$\delta_\mathrm{EW}^{\mathrm{rec}}=\delta_\mathrm{EW}^{\mathrm{bare}}$,
and the results are valid for all three generations.  Hence, the total
\PZ\PH\ cross section with invisible \PZ\ decay can be trivially
obtained by multiplying the given result by three.  On the other hand,
the photon-induced contribution to $\PH \nu_\Pl / \overline{\nu}_\Pl$
depends on the fermion generation and we show numerical results for
the electron channel.  The corrections in the muon and tau channels
are smaller since the corresponding logarithms of the fermion masses
are smaller.

In a single run {\sc Hawk} employs a single set of PDFs and
produces the following NLO prediction,
\beq
\sigma^\mathrm{NLO}_\mathrm{HAWK} 
= \sigma_0 \times \left( 1+ \delta_\mathrm{QCD} + \delta_\gamma 
+ \delta_\mathrm{EW} \right).
\label{eq:sigma_hawk}
\eeq
Note that the LO part $\sigma_0$ is calculated with NLO PDFs in that
case, i.e.\ it is not equal to the proper LO cross section
$\sigma^\mathrm{LO}$, which is to be calculated with LO PDFs and
requires a separate run of {\sc Hawk}.  Likewise,
$(1+\delta_\mathrm{QCD})$ is not equal to the standard QCD $K$-factor
$K_\mathrm{QCD}=\sigma^\mathrm{NLO}_\mathrm{QCD}/\sigma^\mathrm{LO}$.
In the spirit of factorizing QCD and EW corrections, which is
certainly true for corrections that are dominated by soft or collinear
gluons, the fixed-NLO prediction \refeq{eq:sigma_hawk} can be improved
by
\beq
\sigma^\mathrm{NLO}_\mathrm{fact} 
= \sigma^\mathrm{NLO}_\mathrm{QCD} 
\times \left(1 + \delta_\mathrm{EW} \right)
+ \sigma_0\,\delta_\gamma,
\qquad  \sigma^\mathrm{NLO}_\mathrm{QCD} =\si_0\left(1 + \delta_\mathrm{QCD} \right),
\label{eq:sigma_fact}
\eeq
and we show predictions for the electron case, i.e.\
$\delta_\mathrm{EW}=\delta^\mathrm{rec}_\mathrm{EW}$
in the numerical results below.
The NLO EW correction factor $\delta_\mathrm{EW}$ is rather insensitive
to changes in the PDFs, but involves a small dependence on the QED
factorization scale.
Recall that the presently missing ${\cal O}(\alpha)$ corrections to the PDFs
induce an uncertainty at the level $\lsim1\%$, i.e.\ for full
${\cal O}(\alpha)$ precision QED-improved PDFs are needed.
To fix full ${\cal O}(\alpha\alpha_{\mathrm{s}})$ precision beyond this
factorization ansatz, a full two-loop calculation would be required, which
would certainly be overkill for the data analyses at the Tevatron and
LHC. Equation~\refeq{eq:sigma_fact} also offers a simple possibility
to further improve the NLO result to include the NNLO QCD predictions
by other authors~\cite{Brein:2003wg,Ferrera:2011bk} upon replacing
$\sigma^\mathrm{NLO}_\mathrm{QCD}$ by 
$\sigma^\mathrm{NNLO}_\mathrm{QCD}$. For $\PW\PH$ production this is
done in \citere{YR2}, where our results for $\delta_\mathrm{EW}$ are used.

\begin{figure}
\begin{center}
\includegraphics[width=15.3cm]{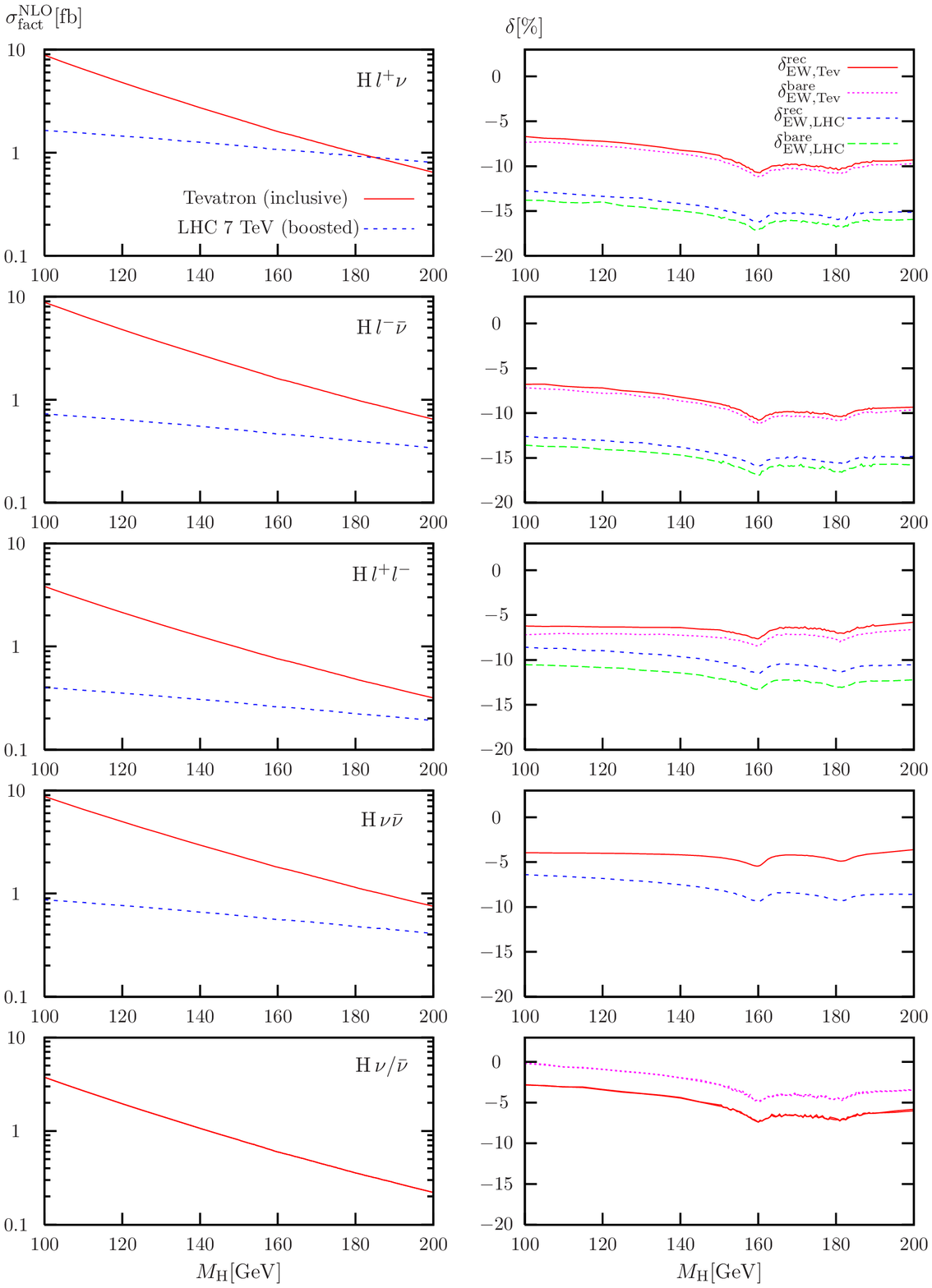}
\caption{\label{fi:mH_plot} The cross section
  $\sigma^\mathrm{NLO}_\mathrm{fact}$ (left)  
  and NLO EW corrections (right) for the different Higgs-strahlung
  processes with basic cuts at the Tevatron as well as for boosted
  Higgs bosons at the $7\TeV$ LHC as a function of the Higgs-boson
  mass.}
\end{center}
\end{figure}

In Figure~\ref{fi:mH_plot} we show our best predictions
$\sigma^\mathrm{NLO}_\mathrm{fact}$ for the cross section and the EW
corrections for the five Higgs-production channels under
consideration. We show the results for the inclusive analysis at the
Tevatron as well as for the boosted analysis at the LHC running at a
CM energy of $7\TeV$. As expected, the EW corrections become more and
more negative with rising CM energy of typical events and, thus, are
larger in absolute terms for the boosted analysis. Their size depends
on the channel and exceeds $-15\%$ for the boosted analysis.
As a function of the Higgs-boson mass, the corrections show small dips
around $\MH=2 \MW$ and $\MH=2 \MZ$ corresponding to threshold
singularities of vector-boson pairs inside loops which are
automatically regularized by the complex-mass scheme in our
calculation. Earlier predictions in \citere{Ciccolini:2003jy} have
been singular at these thresholds. Away from these thresholds, the
corrections for the inclusive analysis differ, as expected, only at
the level of $1{-}2\%$ from the predictions for total cross sections
for on-shell gauge bosons shown in \citere{Ciccolini:2003jy}, mainly
due to the effect of final-state radiation in the presence of lepton
identification cuts.  For the inclusive analysis, the cross section of
the $\PH \nu_\Pl / \overline{\nu}_\Pl$ channel contributes a sizable
fraction to the Higgs plus missing transverse momentum cross section.
For the boosted analysis, where the vector boson has to possess a
large transverse momentum, it is more unlikely for the charged lepton
not to be detected and the contribution due to a missed charged lepton
is negligible (below 0.1 fb at the $7\TeV$ LHC).

\begin{figure}
\includegraphics[width=16cm]{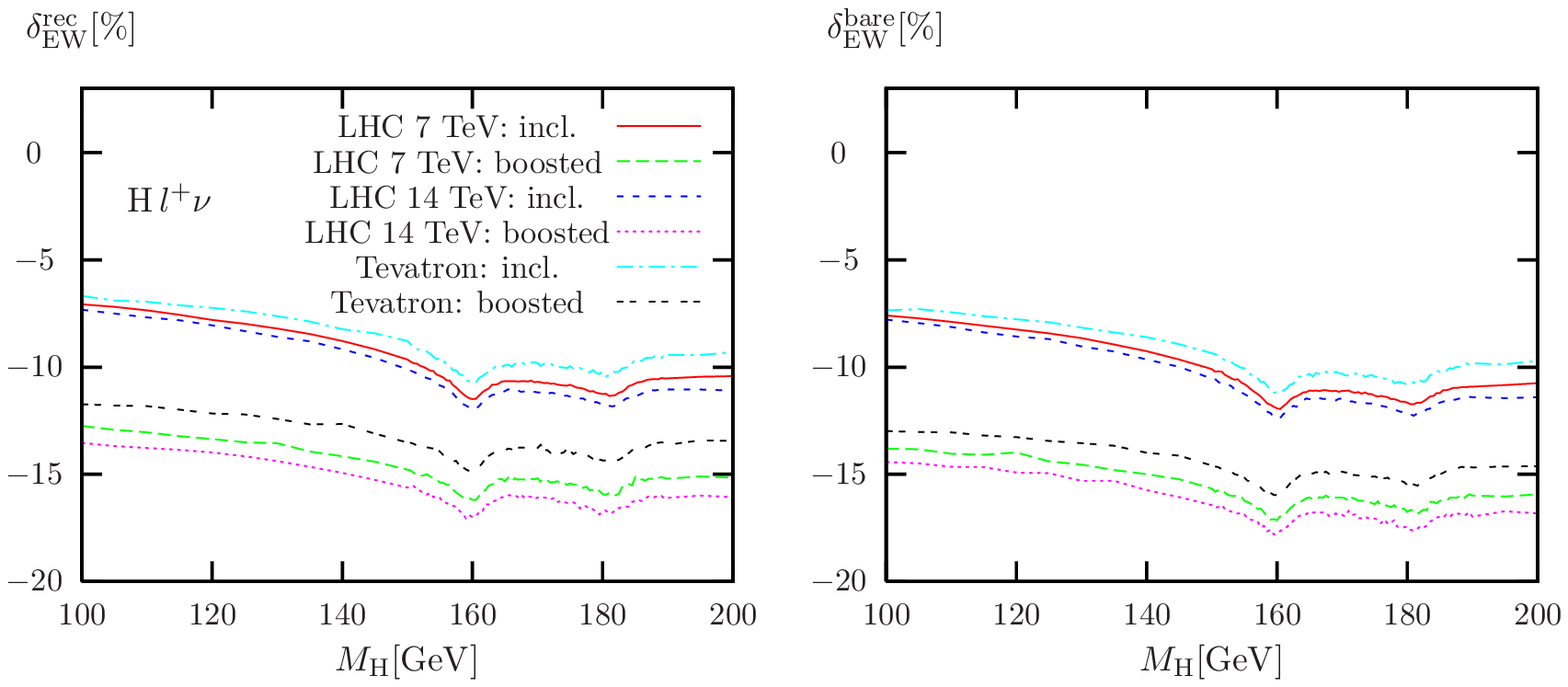}
\caption{\label{fi:compare_colliders} 
 EW corrections with (left) and without (right) lepton--photon
  recombination for $\PH \Pl^+ \nu_\Pl$ production at the LHC at
  $7\TeV$ and at $14\TeV$ and at the Tevatron as a function of the
  Higgs-boson mass.}
\end{figure}

In Figure~\ref{fi:compare_colliders}, we show the EW corrections for
$\PH \Pl^+ \nu_\Pl$ production at the Tevatron and the LHC at $7$ and
$14\TeV$. The differences between the corrections at the colliders are
rather small, at the 1\% level.  As already indicated above, the EW
corrections are insensitive to the choice of the PDF set.  Note that
there is only a small difference of about $1\%$ or $2\%$ between the
cases of bare or recombined leptons, where the larger value of $2\%$
refers to the boosted-Higgs analyses. This small difference reflects
the fact that the major part of the EW correction is not due to
final-state radiation, but due to non-universal weak corrections.

\begin{figure}
\includegraphics[width=8cm]{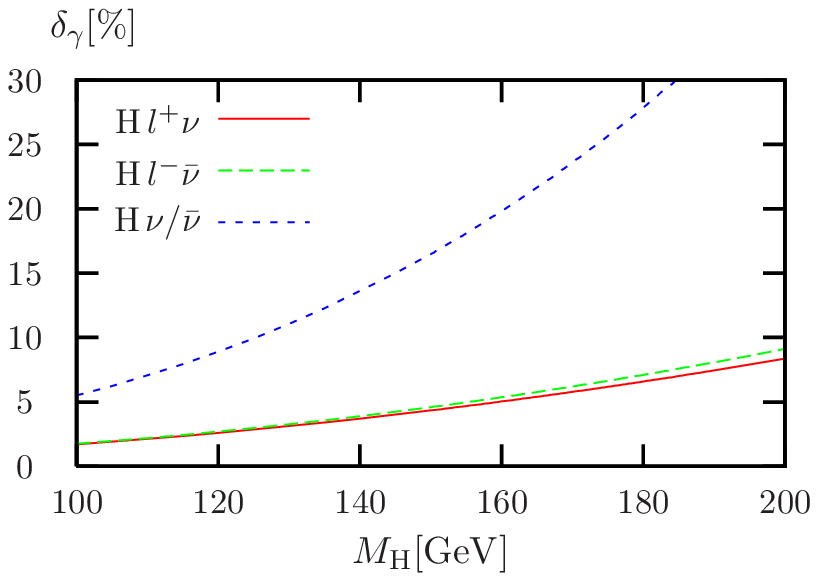}
\includegraphics[width=8cm]{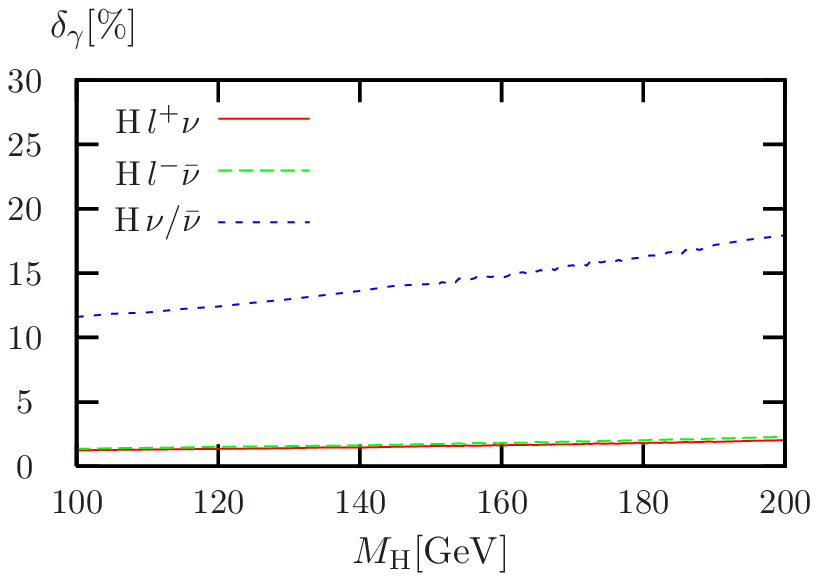}
\caption{\label{fi:gamma_induced} Relative correction from photon-induced processes for 
basic cuts (left) and the boosted-Higgs analysis (right) at the $7\TeV$ LHC
as a function of the Higgs-boson mass.}
\end{figure}

In Figure~\ref{fi:gamma_induced}, we show the contribution from
photon-induced processes to the \PW-mediated channels at the LHC. The
contribution to the channels, where the charged lepton is identified,
is at the $1{-}2\%$ level for the boosted analysis and mainly results from
diagrams where the initial-state photon couples to a $t$-channel \PW\ 
boson. Accordingly, the corrections are almost zero for the
\PZ-mediated channels (not shown).  There are larger photon-induced
corrections for WH production, rising to 8\%, to the inclusive
cross section at the LHC for large Higgs masses around $200\GeV$.
Moreover, the photon-induced corrections to the $\PH \nu_\Pl /
\overline{\nu}_\Pl$ channel, which are additionally enhanced by
logarithms of the charged lepton mass, for boosted Higgs bosons are
larger than 10\% (for the photon splitting into an
$\mathrm{e}^+\mathrm{e}^-$ pair) of the corresponding LO $\PH \nu_\Pl
/ \overline{\nu}_\Pl$ cross section and even more sizable for the
inclusive setup, in particular for high Higgs masses.  
Whether these large corrections become phenomenologically relevant at 
the LHC depends on the details of the analysis.
At the
Tevatron the corresponding corrections do not exceed 3\%.

\begin{table}
$$ \begin{array}{c|c|c|c|c|c}
\mathrm{channel} & \;\PH\Pl^+ \nu_\Pl +X \;& \;\PH\Pl^- \bar\nu_\Pl +X \;& \;\PH\Pl^+ \Pl^- +X \;& \;\PH\nu_\Pl \bar\nu_\Pl +X \;& \;\PH\nu_\Pl / \bar\nu_\Pl +X\; \\ 
  \hline
\sigma_{0}/\fba                                     \; & \; 4.1232(2)        \; & \; 4.1229(2)        \; & \; 1.82773(5)        \; & \; 4.1480(1)          \; & \; 1.6063(2)        \\
\sigma^{\mathrm{LO}}/\fba                           \; & \; 3.6930(5)        \; & \; 3.6926(5)        \; & \; 1.6484(1)         \; & \; 3.7476(4)          \; & \; 1.4355(4)        \\ 
  \hline                                                                                                                 
\delta_{\EW}^{\mathrm{bare}} / \%                   \; & \; -7.8\phz         \; & \; -7.8\phz         \; & \; -7.2\phz          \; & \; -4.1\phz           \; & \; -0.9\phz         \\ 
\delta_{\EW}^{\mathrm{rec}} / \%                    \; & \; -7.3\phz         \; & \; -7.3\phz         \; & \; -6.3\phz          \; & \; -4.1\phz           \; & \; -3.5\phz         \\ 
\delta_{\QCD}/\%                                    \; & \; +24.9\phz        \; & \; +24.9\phz        \; & \; +24.6\phz         \; & \; +24.9\phz          \; & \; +25.1\phz        \\ 
(K_{\mathrm{QCD}}-1)/\%                             \; & \; +39.5\phz        \; & \; +39.5\phz        \; & \; +38.1\phz         \; & \; +38.2 \phz         \; & \; +40.0\phz        \\ 
\delta_{\ga}/\%                                     \; & \; +0.3\phz         \; & \; +0.3\phz         \; & \; +0.0\phz          \; & \; -0.0\phz           \; & \; +1.0\phz         \\ 
  \hline                                                                                                                 
\sigma^{\mathrm{NLO}}_{\mathrm{fact}}/\fba          \; & \; 4.7884(5)        \; & \; 4.7872(5)        \; & \; 2.1332(1)         \; & \; 4.9696(3)          \; & \; 1.9566(4)        \\ 
\sigma^{\mathrm{NLO}}_{\mathrm{HAWK}}/\fba          \; & \; 4.8635(5)        \; & \; 4.8622(5)        \; & \; 2.1616(1)         \; & \; 5.0115(3)          \; & \; 1.9706(4)        \\ 
\end{array} $$

\vspace{-0.5cm}
\caption{\label{ta:incl} Born cross sections $\sigma_0$ and 
$\sigma^\mathrm{LO}$ evaluated with NLO and LO PDFs, respectively, 
various relative corrections, and the resulting predictions 
for inclusive production in 
the various Higgs-strahlung processes ($\MH=120\GeV$, Tevatron)
according to \refeq{eq:sigma_hawk} and \refeq{eq:sigma_fact}. All relative
corrections $\delta$ are given relative to $\sigma_0$, and $K_\mathrm{QCD}
=\sigma^\mathrm{NLO}_{\mathrm{QCD}}/\sigma^\mathrm{LO}$.}
\end{table}
\begin{table}
$$ \begin{array}{c|c|c|c|c|c}
\mathrm{channel} & \;\PH\Pl^+ \nu_\Pl +X \;& \;\PH\Pl^- \bar\nu_\Pl +X \;& \;\PH\Pl^+ \Pl^- +X \;& \;\PH\nu_\Pl \bar\nu_\Pl +X \;& \;\PH\nu_\Pl / \bar\nu_\Pl +X\; \\ 
  \hline
\sigma_{0}/\fba                                     \; & \; 1.50846(7)       \; & \; 0.66292(3)       \; & \; 0.35349(2)       \; & \; 0.74759(3)       \; & \; 0.058236(9)     \\ 
\sigma^{\mathrm{LO}}/\fba                           \; & \; 1.4183(2)        \; & \; 0.60926(9)       \; & \; 0.32845(5)       \; & \; 0.69519(9)       \; & \; 0.05417(3)      \\ 
  \hline                                                                                                                 
\delta_{\EW}^{\mathrm{bare}} / \%                   \; & \; -14.2\phz        \; & \; -14.0\phz        \; & \; -10.9\phz        \; & \; -6.9\phz         \; & \; -12.5\phz       \\ 
\delta_{\EW}^{\mathrm{rec}} / \%                    \; & \; -13.3\phz        \; & \; -13.0\phz        \; & \; -9.0\phz         \; & \; -6.9\phz         \; & \; -14.5\phz    \\ 
\delta_{\QCD}/\%                                    \; & \; +9.5\phz         \; & \; +9.4\phz         \; & \; +9.8\phz         \; & \; +9.8\phz         \; & \; +6.8\phz     \\ 
(K_{\mathrm{QCD}}-1)/\%                             \; & \; +16.5\phz        \; & \; +19.1\phz        \; & \; +18.1\phz        \; & \; +18.1\phz        \; & \; +14.9\phz    \\ 
\delta_{\ga}/\%                                     \; & \; +1.3\phz         \; & \; +1.5\phz         \; & \; +0.0\phz         \; & \; +0.0\phz         \; & \; +12.5\phz    \\ 
  \hline                                                                                                                 
\sigma^{\mathrm{NLO}}_{\mathrm{fact}}/\fba          \; & \; 1.4522(4)        \; & \; 0.6406(2)        \; & \; 0.35329(7)       \; & \; 0.7646(2)       \; & \; 0.06043(6)       \\
\sigma^{\mathrm{NLO}}_{\mathrm{HAWK}}/\fba          \; & \; 1.4713(4)        \; & \; 0.6488(2)        \; & \; 0.35639(7)       \; & \; 0.7697(2)       \; & \; 0.06100(6)       \\ 
\end{array} $$

\vspace{-0.5cm}
\caption{\label{ta:boosted} Born cross sections, various corrections, and the
resulting predictions for boosted production in the various 
Higgs-strahlung processes ($\MH=120\GeV$, LHC at $7\TeV$). See \refta{ta:incl}
for details.}
\end{table}

To facilitate a possible comparison of our results with future results of
other groups, in Tables~\ref{ta:incl} and \ref{ta:boosted}, 
we show the cross sections and corrections for $\MH=120$~GeV. 

\begin{figure}
\includegraphics[width=7.9cm, bb = 143 74 389 724]{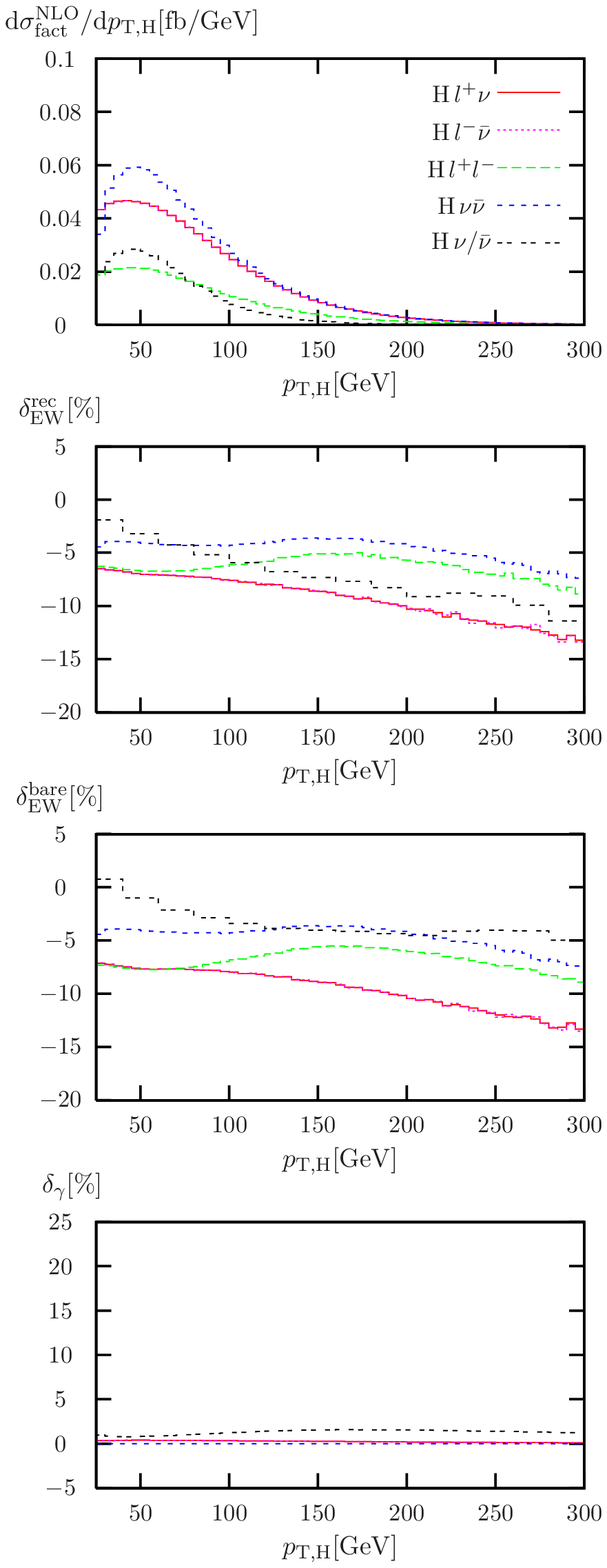}
\includegraphics[width=7.9cm, bb = 143 74 389 724]{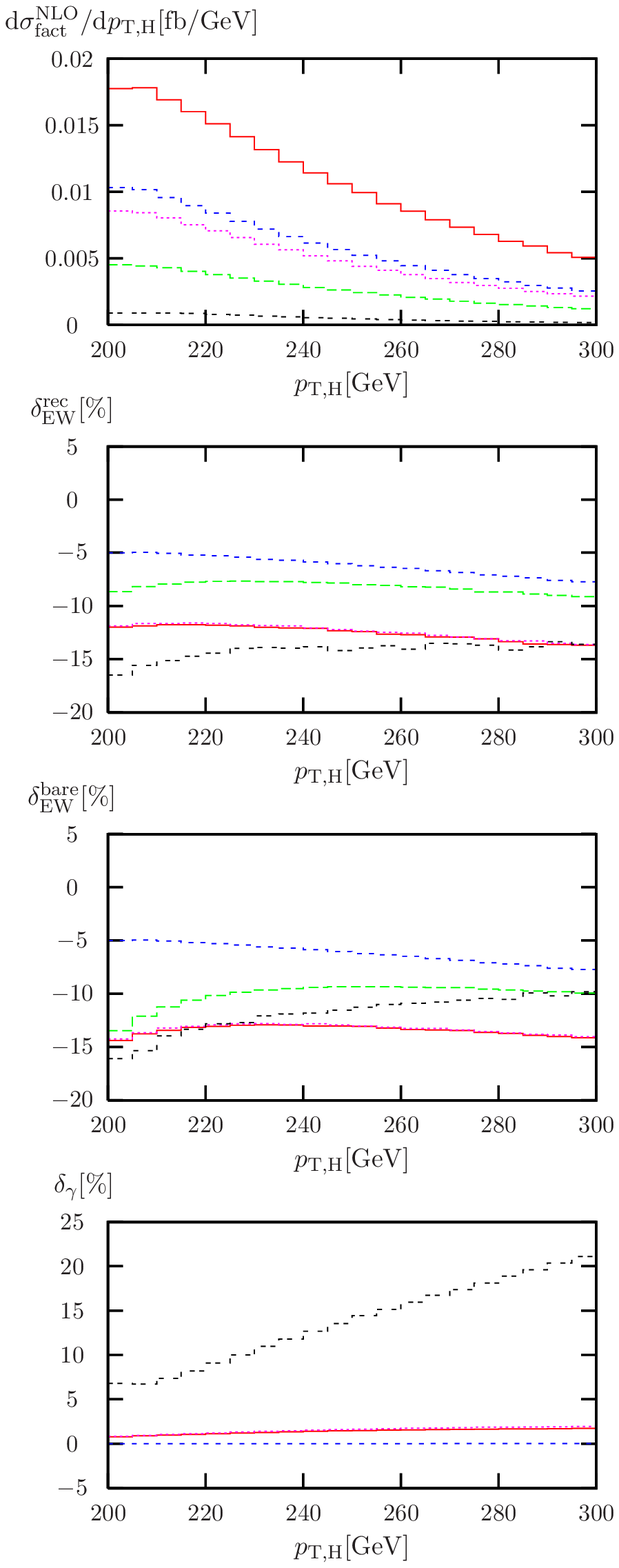}
\caption{\label{fi:pth}
  The $p_{\mathrm{T},\PH}$ distributions
  $\mathrm{d}\sigma^\mathrm{NLO}_\mathrm{fact}/\mathrm{d}p_{\mathrm{T},\PH}$
  (top), NLO EW corrections for recombined and bare leptons (middle),
  and corrections due to photon-induced processes (bottom) for
  Higgs-strahlung with basic cuts at the Tevatron (left) as well as
  for boosted Higgs bosons at the $7\TeV$ LHC (right) for $\MH=120\GeV$.}
\end{figure}

Figure~\ref{fi:pth} shows the differential cross sections with respect
to the transverse momentum of the Higgs boson up to $300\GeV$, along
with the EW corrections. For the inclusive Higgs-boson sample, the EW
corrections range between $-5\%$ and $-15\%$ and are largest for the
\PW-mediated channels at large $p_\mathrm{T,\PH}$.  The difference due
to the treatment of final-state photons is again small, since the bulk
of the correction is of weak origin and not due to final-state
radiation. The same observation holds for large values of the
transverse momentum in the boosted-Higgs analysis.  However, close to
the cut value of $200\GeV$ for $p_\mathrm{T,\PH}$, the corrections
receive a sizeable negative contribution from final-state radiation
effects, in particular for the $\PH\Pl^+\Pl^-$ channel, since we also
require $p_{\mathrm{T},V} > 190\GeV$.  Final-state radiation can
destroy the perfect balance between the Higgs- and the vector-boson
transverse momentum at Born level. This balance is also influenced by
initial-state radiation and would lead to a breakdown of the
fixed-order calculation near the cut if the cuts on $p_{\mathrm{T,H}}$
and $p_{\mathrm{T,W/Z}}$ were equal.  As mentioned above, the
photon-induced contributions are large for the 
$\PH \nu_\Pl/\overline{\nu}_\Pl$ channel in the boosted analysis, but
the overall contribution is small and phenomenologically unimportant.

\begin{figure}
\includegraphics[width=7.9cm, bb = 143 74 389 724]{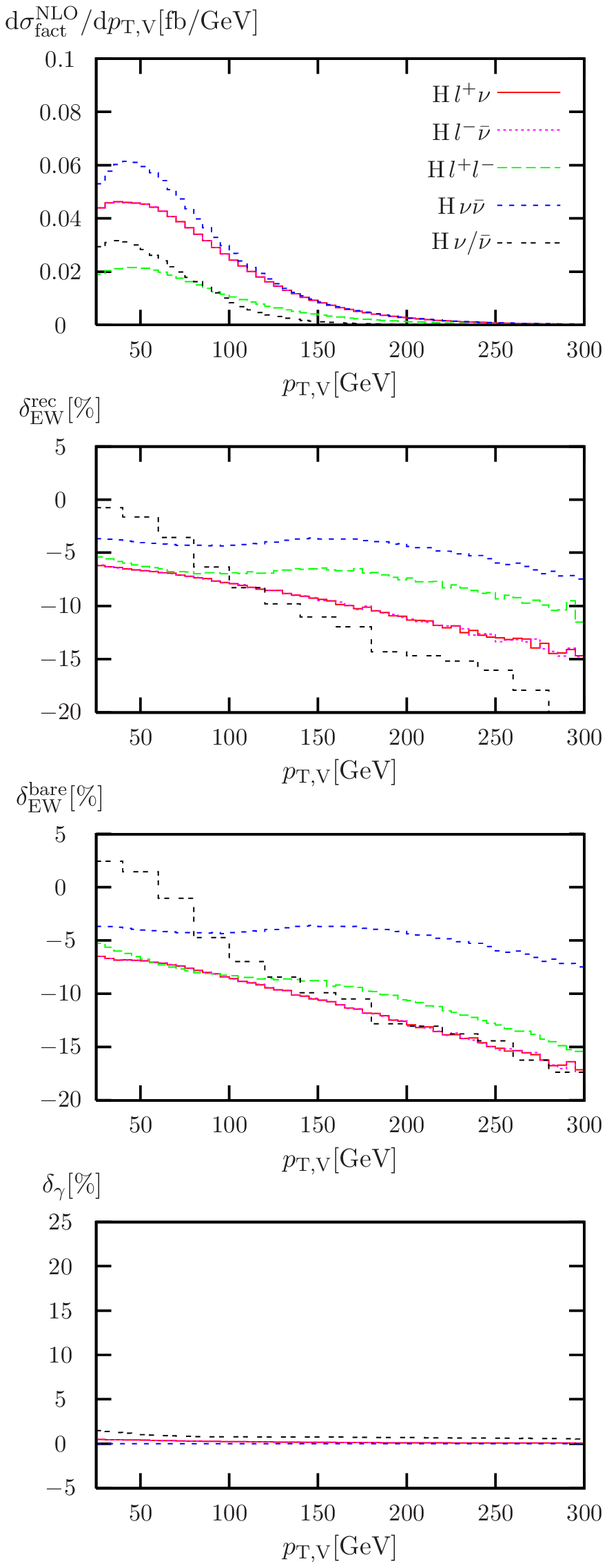}
\includegraphics[width=7.9cm, bb = 143 74 389 724]{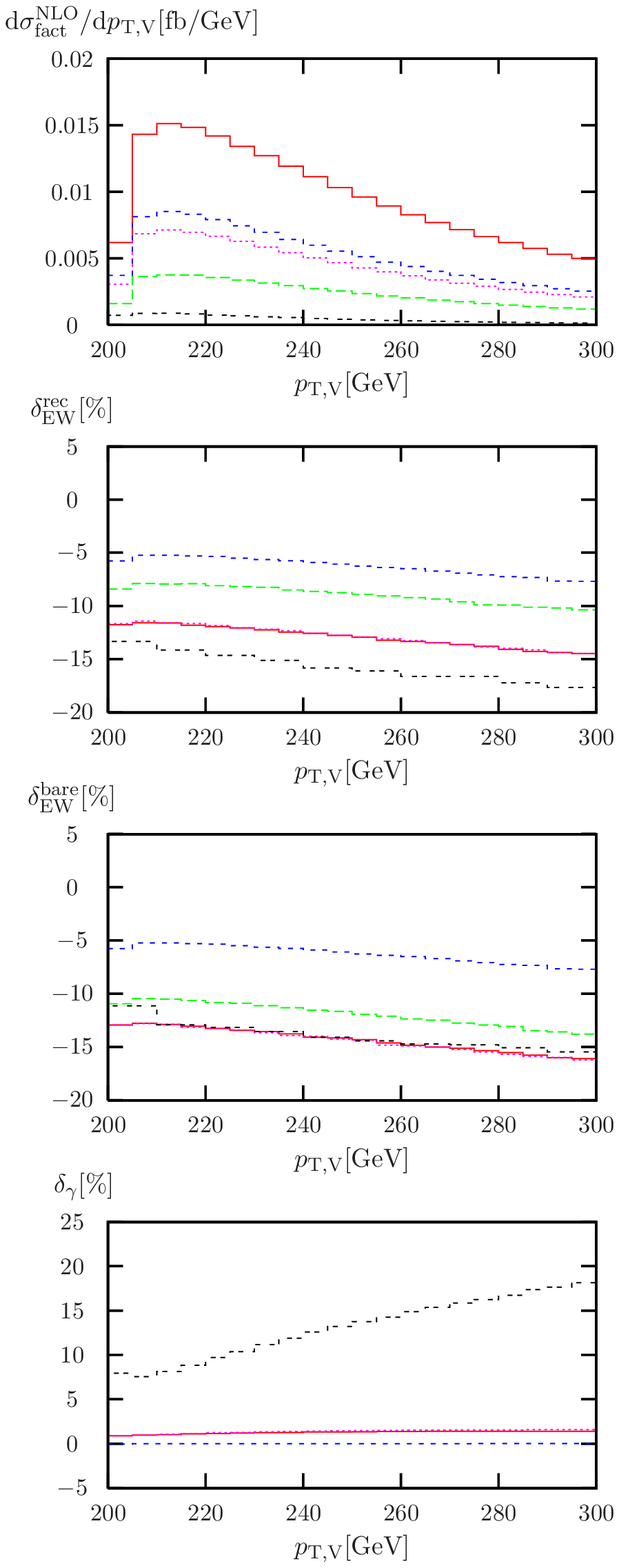}
\caption{\label{fi:ptV}
  The $p_{\mathrm{T},V}$ distributions
  $\mathrm{d}\sigma^\mathrm{NLO}_\mathrm{fact}/\mathrm{d}p_{\mathrm{T},V}$
  (top), NLO EW corrections for recombined and bare leptons (middle),
  and corrections due to photon-induced processes (bottom) for
  Higgs-strahlung with basic cuts at the Tevatron (left) as well as
  for boosted Higgs bosons at the $7\TeV$ LHC (right) for
  $\MH=120\GeV$.}
\end{figure}

Figure~\ref{fi:ptV} shows the differential cross section with respect
to the transverse momentum of the vector boson, i.e. the sum of the
transverse momenta of the leptons in the final state (or the transverse 
momentum of the neutrino in the $\PH \nu_\Pl / \overline{\nu}_\Pl$ case, 
as discussed at the end of \refse{se:calc}). The EW
corrections are very similar to the ones for the $p_\mathrm{T,\PH}$
distribution. Only the large contribution due to final-state radiation
near the cut in the boosted analysis is absent because the Higgs boson
is uncharged (possible final-state radiation of the Higgs decay
products is not available in our analysis).  However, near
$p_{\mathrm{T,W/Z}} = 200$~GeV the perturbative uncertainty due to QCD
initial-state radiation rises, since there is no tree-level
distribution below the value of the $p_{\mathrm{T,H}}$ cut.  In
Figure~\ref{fi:ptV}, one observes a sharp drop of the cross section in
the first bin of the distribution which is due to correspondingly large
NLO QCD corrections.

\begin{figure}
\includegraphics[width=7.9cm, bb = 143 74 389 724]{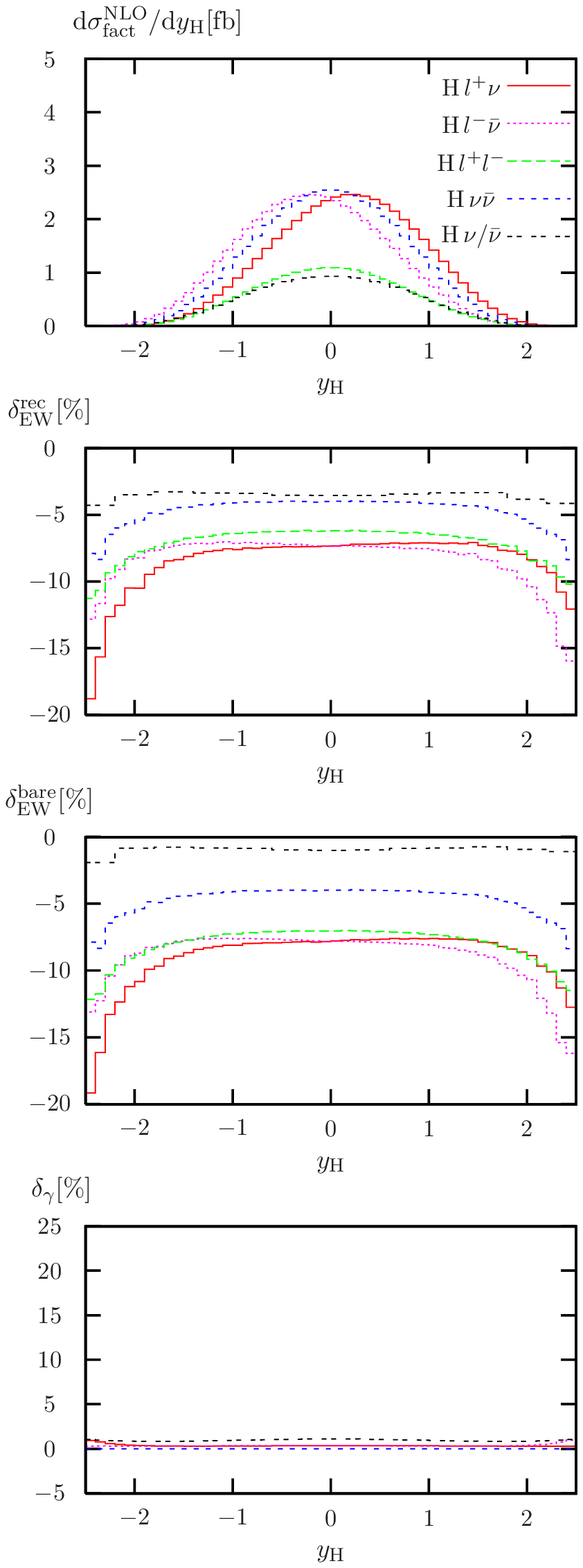}
\includegraphics[width=7.9cm, bb = 143 74 389 724]{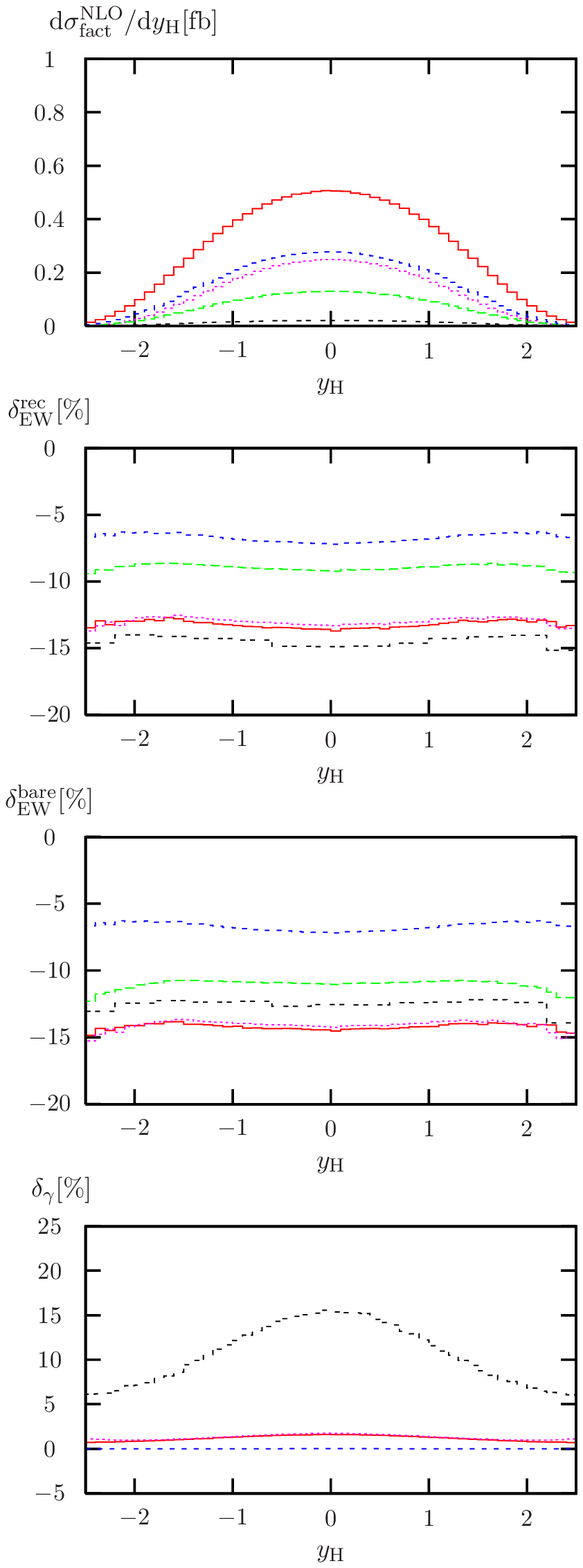}
\caption{\label{fi:yh}
The $y_{\PH}$ distributions
$\mathrm{d}\sigma^\mathrm{NLO}_\mathrm{fact}/\mathrm{d}y_{\PH}$
(top), NLO EW corrections for recombined 
and bare leptons (middle), and corrections due to photon-induced processes
(bottom) for Higgs-strahlung
with basic cuts at the Tevatron (left) as well as for boosted Higgs bosons at the $7\TeV$ LHC (right)
for $\MH=120\GeV$.}
\end{figure}

\begin{figure}
\includegraphics[width=7.9cm, bb = 143 74 389 724]{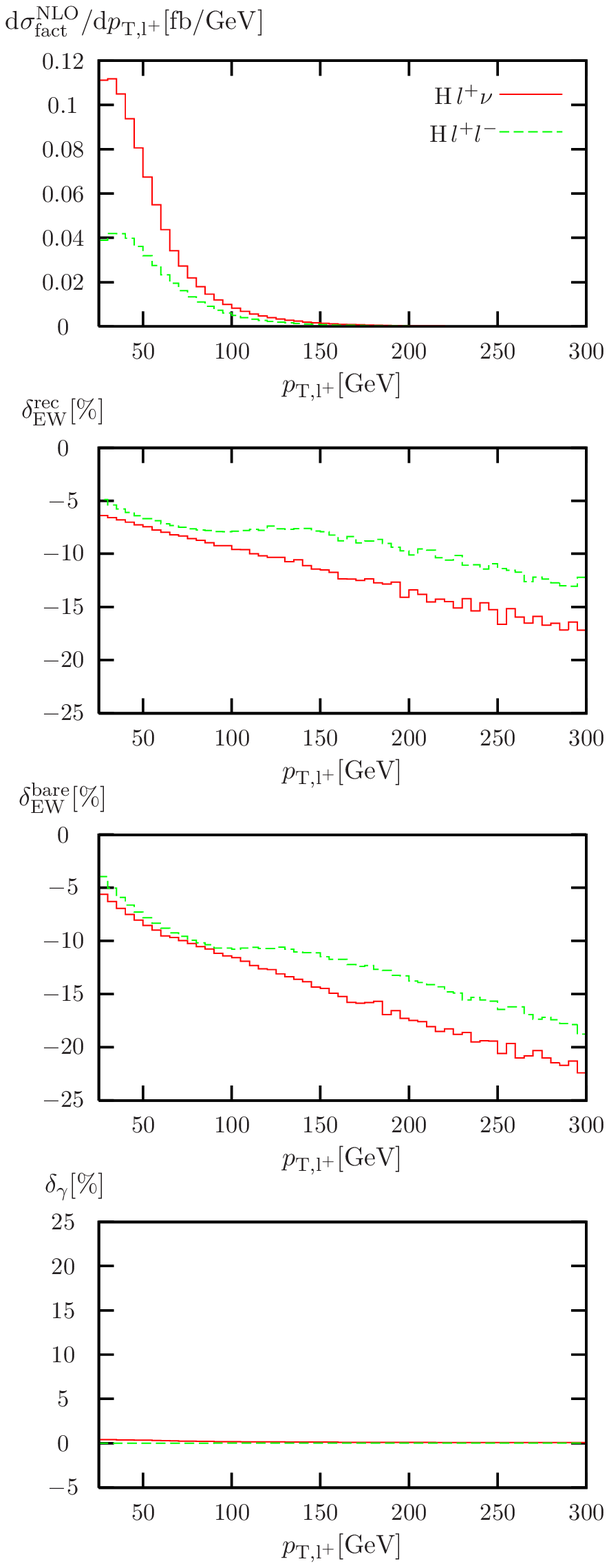}
\includegraphics[width=7.9cm, bb = 143 74 389 724]{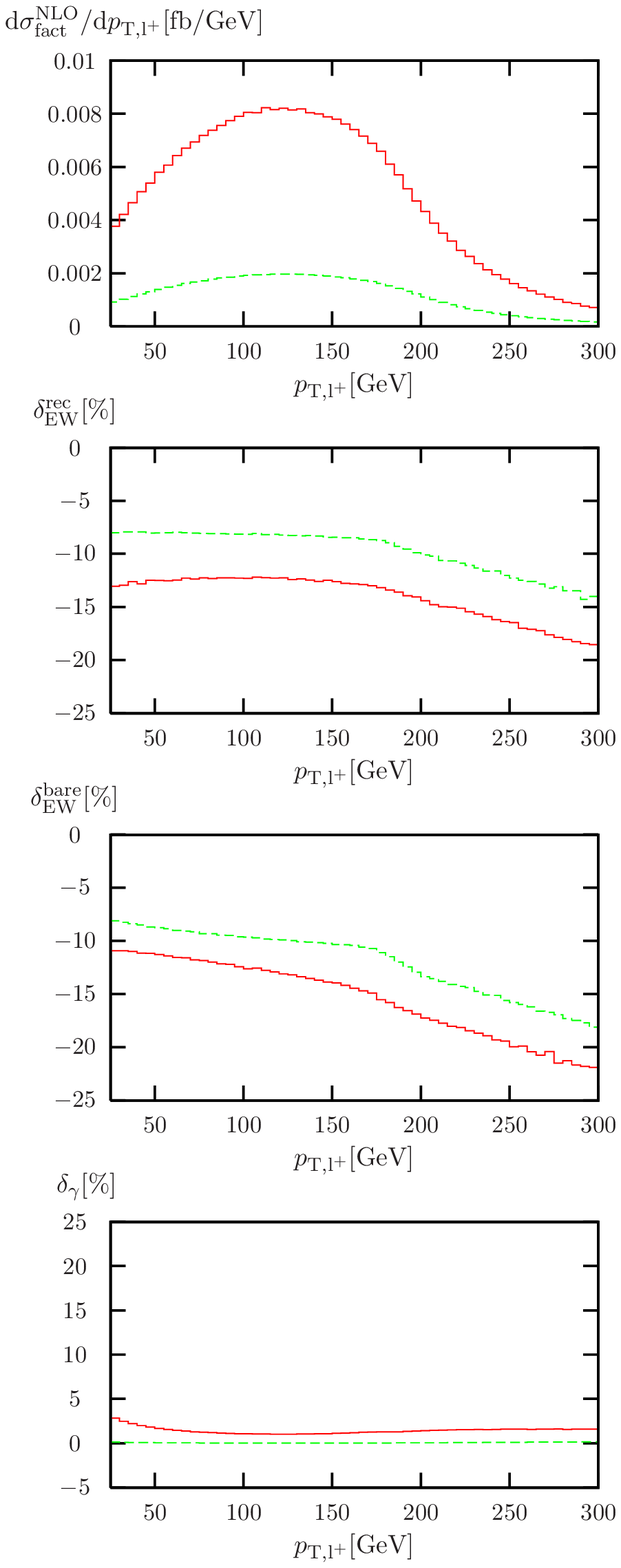}
\caption{\label{fi:ptl}
The $p_{\mathrm{T},l^+}$ distributions
$\mathrm{d}\sigma^\mathrm{NLO}_\mathrm{fact}/\mathrm{d}p_{\mathrm{T},l^+}$
(top), NLO EW corrections for recombined 
and bare leptons (middle), and corrections due to photon-induced processes
(bottom) for Higgs-strahlung
with basic cuts at the Tevatron (left) as well as for boosted Higgs bosons at the $7\TeV$ LHC (right)
for $\MH=120\GeV$.}
\end{figure}

The differential distributions with respect to the rapidity of the
Higgs boson and the transverse momentum of a positively charged lepton
in the final state are displayed in Figures~\ref{fi:yh} and
\ref{fi:ptl}, respectively. The EW corrections for the rapidity
distribution are almost completely flat and, hence, resemble the
correction for the cross section using a given set of cuts. The
transverse-momentum distribution shows corrections growing in absolute
size with the energy of the typical events as observed before, close
to or even beyond $-20\%$ at $p_{\mathrm{T},\Pl^+} = 300\GeV$ with
recombination or for bare muons, respectively.

\begin{figure}
\includegraphics[width=7.9cm, bb = 143 74 389 724]{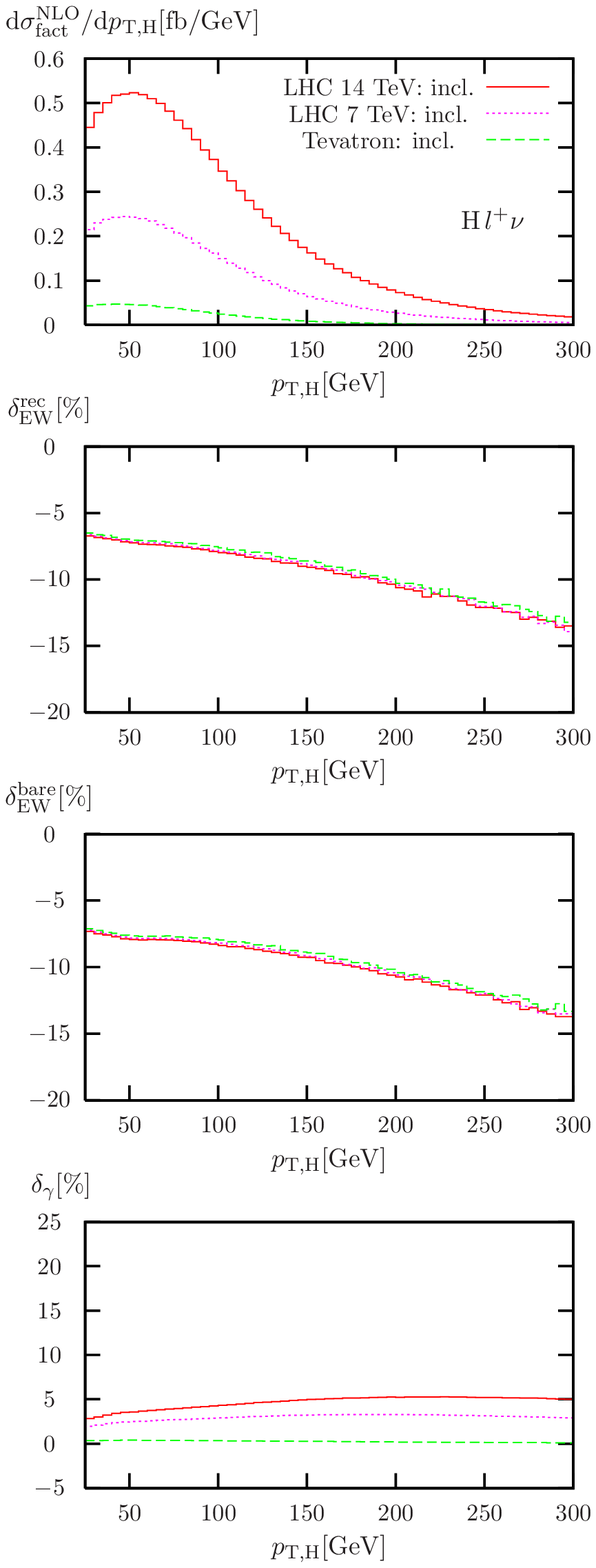}
\includegraphics[width=7.9cm, bb = 143 74 389 724]{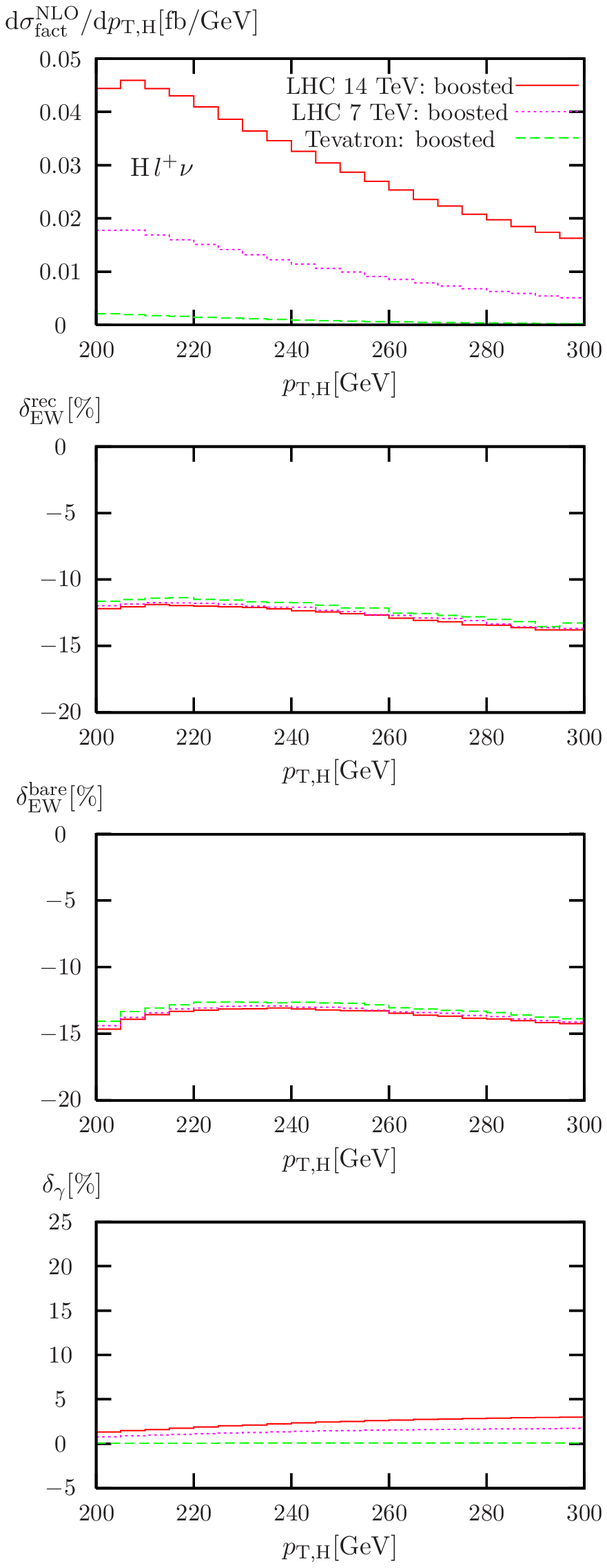}
\caption{\label{fi:pth_compare_colliders}
  The $p_{\mathrm{T},\PH}$ distributions
  $\mathrm{d}\sigma^\mathrm{NLO}_\mathrm{fact}/\mathrm{d}p_{\mathrm{T},\PH}$
  (top), NLO EW corrections for recombined and bare leptons (middle)
  and corrections due to photon-induced processes (bottom) for for
  $\PH \Pl^+ \nu_\Pl$ production and different collider setups with
  basic cuts (left) and additional cuts for boosted-Higgs analyses
  (right) for $\MH=120\GeV$.}
\end{figure}

Finally, Figure~\ref{fi:pth_compare_colliders} shows
that for the $\PH\Pl^+ \nu_\Pl$ channel the EW corrections hardly
differ for the same observable measured at different hadron colliders.
Only the photon-induced corrections are more sensitive to the specific
collider settings.  Exemplarily, the distribution with respect to the
transverse momentum of the Higgs boson is shown but other
distributions show the same qualitative behaviour.

\section{Conclusions}
\label{se:conclusion}

In this paper we have discussed the impact of electroweak
radiative corrections to the Higgs-strahlung processes
off W/Z bosons,
$\Pp\Pp/\Pp\bar\Pp\to\PH l\nu_l/l^-l^+/\nu_l\bar\nu_l+X$,
at the Tevatron and the LHC.
Compared to a previous calculation, we provide fully differential 
predictions. We include the leptonic decays of
the W and Z bosons and support the full kinematical information
on the Higgs boson and the decay leptons of the weak gauge bosons.
These processes and the corresponding next-to-leading-order QCD and 
electroweak corrections are included in the new version 2.0 of
the Monte Carlo program {\sc Hawk},
which was initially designed for 
Higgs + 2 jet production via
vector-boson fusion and 
WH/ZH production with hadronically decaying W/Z
bosons.

The electroweak corrections are of the order of $-(5{-}10)\%$ for
total cross sections and, thus, larger than the uncertainty
originating from parton distribution functions and QCD factorization
and renormalization scales.  Our detailed discussion of the
electroweak corrections to transverse-momentum and rapidity
distributions of the Higgs boson and outgoing charged leptons
demonstrates that the corrections further increase in size in the
regions of increasing transverse momenta.  For instance, imposing the
cut $p_{\mathrm{T,H}}>200\GeV$ at the LHC, which is part of the search
strategy for highly-boosted Higgs bosons decaying into ``fat jets''
with $\Pb\bar\Pb$ substructure, drives the corrections to WH
production to about $-14\%$ for $\MH=120\GeV$.  The electroweak
corrections to the different $p_{\mathrm{T}}$ distributions can
already exceed $-15\%$ in the range $p_{\mathrm{T}} = 200{-}300\GeV$
and grow further when more energetic events are analyzed.

Electroweak corrections, thus, represent an important ingredient
in the theory predictions for Higgs-strahlung off W/Z bosons
at the Tevatron and the LHC. 

\subsection*{Acknowledgements}

This work is supported in part by the Gottfried Wilhelm 
Leibniz programme of the Deutsche Forschungsgemeinschaft (DFG).
We thank the members of the LHC Higgs Cross Section Working Group, in
particular J. Olsen and G. Piacquadio for useful discussions on
the experimental setup.

\end{document}